\documentclass[12pt]{article}
\pdfoutput=1
\usepackage{bm}
\usepackage{amsmath}
\usepackage{amsfonts}
\numberwithin{equation}{section}
\usepackage{amssymb}
\usepackage{graphicx}
\usepackage{bm}
\usepackage{float}
\usepackage{cite}
\usepackage{bbold}
\usepackage{multirow}
\usepackage{enumerate}
\usepackage{dynkin-diagrams}
\usepackage{ulem}
\usepackage{color}

\usepackage[utf8]{inputenc}
\usepackage[hidelinks]{hyperref}
\begin{document}
\begin{titlepage}
		\begin{flushright}
			TIT/HEP-694 \\
			May 2023
		\end{flushright}
		\vspace{0.5cm}
		\begin{center}
			{\Large \bf
				WKB analysis of the linear problem for modified affine Toda field equations
			}
			\lineskip.75em
			\vskip 2.5cm
			{\large  Katsushi Ito, Mingshuo Zhu }
			\vskip 2.5em
			{\normalsize\it Department of Physics,\\
				Tokyo Institute of Technology\\
				Tokyo, 152-8551, Japan}
			\vskip 3.0em
		\end{center}
		\begin{abstract}
We study the WKB analysis of the solutions to the linear problem for a modified affine Toda field equation,
which is equivalent to the higher-order ordinary differential equation (ODE) studied in the ODE/IM correspondence.
After gauge transformation, we diagonalize the flat connection of the linear problem to reduce the latter to a set of independent first-order linear differential equations. 
We explicitly perform this procedure for classical affine Lie algebras with lower ranks. 
In particular, we study the WKB solutions of the $D_r^{(1)}$- and $D^{(2)}_{r+1}$-type linear problems, which correspond to the higher-order ODEs with the pseudo-differential operator.
The diagonalized connection is obtained from the Riccati equations of the adjoint linear problem and related to the conserved currents of the integrable hierarchy constructed by Drinfeld and Sokolov up to total derivatives.

		\end{abstract}
\end{titlepage}

	\baselineskip=0.7cm
 	\numberwithin{equation}{section}
	\numberwithin{figure}{section}
	\numberwithin{table}{section}

\section{Introduction}
The exact WKB analysis based on the Borel resummation \cite{AIHPA_1983__39_3_211_0} of the second-order ordinary differential equations has been studied extensively in the past decades (see \cite{KaTa} for review).
It also plays a vital role in the research of the ODE/IM correspondence,
which relates the spectral analysis of the ODE to the functional relation approach in the quantum integrable models (IMs) \cite{Dorey:1999uk}.
The connection coefficients between the asymptotic solutions of the ODE around the singularities define the Q-, T-, and Y-functions of the quantum integrable model \cite{Kuniba:2010ir}. Their consistency conditions and analytical properties determine the functional relations among them.
The exact WKB periods correspond to the logarithm of the Y-functions in the quantum integrable models, which has been used to reformulate the spectral problem of the Schr\"odinger equation in terms of the solution to the TBA equations \cite{Ito:2018eon}. In the context of supersymmetric gauge theories, see \cite{Gaiotto:2009hg}.

The exact WKB method for higher-order ODEs is a new research subject since 
\cite{doi:10.1063/1.525467,aoki1994,aoki1998exact}. It becomes important in different contexts: the quantum Seiberg-Witten curves in supersymmetric gauge theories \cite{Mironov:2009uv,Basar:2015xna,Kashani-Poor:2015pca,Ito:2017ypt,Ito:2018hwp,Grassi:2019coc,Ito:2019twh,Ito:2020lyu} and the AdS/CFT correspondence \cite{Alday:2009dv,Alday:2010vh,Hatsuda:2010cc}. Another example is the ODE/IM correspondence between the higher order ODEs and the quantum integrable model associated with affine Lie algebras, where the Q-functions and the T-Q relations have been mostly studied \cite{Dorey:1999pv,Suzuki:1999hu,Dorey:2000ma,Dorey:2006an,Sun:2012xw, Ito:2013aea,Masoero:2015lga,Ito:2020htm,Ekhammar:2020enr}.
In \cite{Ito:2021boh, Ito:2021sjo}, the WKB analysis of the higher-order ODEs and their relation to the TBA system has been investigated. It is observed that the minimal chamber, the $(A_2,A_r)$-type TBA system reproduces the asymptotic expansion of the WKB periods. An interesting phenomenon observed in \cite{Ito:2021boh} is that some specific terms become total derivatives in the WKB expansion of the solutions.
It can be shown that, for the Schr\"odinger equation, the odd terms in $\hbar$ become the total derivatives. For $(r+1)$-th order ODE with even $r$, some even terms also become the total derivative. It results in the absence of the corresponding terms in the Y-functions, which is also confirmed by the TBA equations.

The Schr\"odinger equation can be expressed as a rank-two first-order linear differential system, so one can apply the WKB analysis in the second-order ODE to the rank-two linear system.
The relevant linear system is also obtained from the light-cone limit of the linear problem associated with the $A_1^{(1)}$ affine Toda field equation (the Sinh-Gordon equation) modified by the conformal transformation. The linear problem associated with the affine Toda field equations appears in the massive ODE/IM correspondence \cite{Lukyanov:2010rn, Ito:2013aea,Adamopoulou:2014fca,Ito:2018wgj}, where Baxter's T-Q relations are related to the connection coefficients between the solutions at the origin and infinity. 
To study the Y-system and the TBA system through the ODE/IM correspondence, it is necessary to develop the WKB analysis of the linear problem for the modified affine Toda field equations or related higher-order ordinary differential equations.  
Meanwhile, the quantum Seiberg-Witten curve takes the form of the Hitchin system, a set of first-order differential equations on the Riemann surface, which can be reduced to a higher-order ODE. After applying the WKB method and the functional relation approach to the quantum integrable models, it is possible to find the quantum SW periods which solve the theory in the NS limit of the Omega background \cite{Fioravanti:2019awr,Yan:2020kkb}.  
The procedure to obtain the higher-order ODE is called "abelianization", which relates the gauge bundle on the Riemann surface to the line bundle \cite{Hollands:2019wbr,Yan:2020kkb,Dumas:2020zoz}.

The purpose of this paper is to study the WKB analysis of the linear problem associated with the modified affine Toda field equation for an affine Lie algebra $\hat{\mathfrak g}$. For $\hat{\mathfrak{g}}=A_r^{(1)}$, $A_{2r-1}^{(2)}$, $D^{(2)}_{r+1}$, and $D_r^{(1)}$, the corresponding ODEs, which lead to the Bethe ansatz equations for the Langlands dual $A_r^{(1)}$,  $B^{(1)}_r$, $C_r^{(1)}$, and $D_r^{(1)}$, have been given in \cite{Dorey:2006an, Ito:2013aea}.
However, the ODEs for the D-types and some other exceptional affine Lie algebras contain the pseudo-differential operator $\partial^{-1}$ making the ODE's WKB analysis difficult to study.

This work explores the WKB analysis of the linear system after diagonalizing the matrix-valued Lax operator. It is equivalent to directly conducting the WKB analysis of the higher-order ODEs with the same affine Lie algebra construction. Moreover, our diagonalization approach is closely related to the Drinfeld-Sokolov structure of infinite conserved charges in the generalized KdV hierarchies \cite{Drinfeld:1984qv} since our 
linear differential operators belong to the class of the Lax operators considered there.
We will see that the WKB expansion of the solution can be obtained from the conformal transformation of the classical conserved densities. 
It also predicts the order of the non-trivial terms in the WKB periods from the viewpoint of the classical conserved charges in the hierarchies.
We will apply the diagonalization method to affine Lie algebras $A_r^{(1)}$, $A_{2r-1}^{(2)}$, $D^{(2)}_{r+1}$, $D_r^{(1)}$, and $B^{(1)}_r$. For lower-rank cases, we can explicitly confirm the equivalence between the WKB expansions and the classical conserved charges. 

This paper is organized as follows. In Section \ref{Sec 2}, we review the linear problems for affine Toda field equations and their WKB expansion with the direct method.  In Section \ref{sect:Sec3}, we provide the diagonalization approach and apply this method to the linear problem for $A_{1}^{(1)}$, $A_{2}^{(1)}$, and generalize it to $A_{r}^{(1)}$.  In Section \ref{sect:Sec_4}, we generalize the diagonalization method to other affine Lie algebra types and show that the diagonalization for $A_{2r-1}^{(2)}$, $B^{(1)}_r$, $D^{(2)}_{r+1}$, and $D_r^{(1)}$. In Section \ref{Sec_5}, We find the diagonal elements are all conserved densities of the KdV hierarchies. In Appendix \ref{Appe.A}, we give the representations used in this paper. In Appendix \ref{Appe.B}, we provide a matrix WKB method to solve $D_{r+1}^{(2)}$-type linear problems. Finally, we list the higher-order corrections of the diagonal elements in a specific limit in Appendix \ref{Appe.c}.

\section{The linear problem associated with the modified affine Toda field equations and its WKB analysis}\label{Sec 2}
In this section, we introduce the linear problem for modified affine Toda field equations and show how they relate to higher-order ODEs. Before that, let us first fix the notation for Lie algebras following \cite{Kac:1990gs}.

Let $\mathfrak{g}$ be a simple Lie algebra of rank $r$.
We denote $\alpha_i$ $(i=1,\dots, r)$ as the simple roots and $E_{\alpha_{i}}$, $E_{-\alpha_{i}}$, $H_{i}=\alpha^{\vee}_i\cdot H$ ($i=1,\dots, r$) as the Chevalley generators of $\mathfrak{g}$ satisfying the commutation relations:
\begin{align}\label{eq:superalgebra}
\begin{split}
    &[E_{\alpha_{i}}, E_{-\alpha_{i}}]=H_{i},\\ [H_{i}, E_{\alpha_{j}}]=a_{ij}&E_{\alpha_{j}},\quad [H_{i},E_{-\alpha_{j}}]=-a_{ij}E_{-\alpha_{j}}.\\
    \end{split}
\end{align}
$\alpha^{\vee}$ denotes the coroot of $\alpha$ with $\alpha^{\vee}=2\alpha/|\alpha|^{2}$.
The Cartan matrix $a_{ij}$ is defined by
\begin{equation}
    a_{ij}=\frac{2\alpha_{i}\cdot\alpha_{j}}{\alpha_{j}^{2}}=\alpha_{i}\cdot\alpha_{j}^{\vee}.
\end{equation}
The (co)fundamental weights $\omega_{i}(\omega_{i}^{\vee})$ are the vectors dual to $\alpha_{i}(\alpha_{i}^{\vee})$ satisfying
\begin{equation}
\omega_{i}\cdot\alpha_{j}^{\vee}=\omega_{i}^{\vee}\cdot\alpha_{j}=\delta_{ij},\quad (1\leq i\leq r)
\end{equation}
and the (co)Weyl vectors $\rho(\rho^{\vee})$ is the sum of the (co)fundamental weights. 
The simple Lie algebra $\mathfrak{g}$ is equipped with the canonical gradation:
\begin{equation}\label{eq: can grad}
    \mathfrak{g}=\bigoplus_{k\in\mathbf{Z}}\mathfrak{g}^{k},\quad [\mathfrak{g}^{k},\mathfrak{g}^{l}]\subset{\mathfrak{g}^{k+l}}
\end{equation}
with degree $+1(-1)$ to $E_{\alpha_{i}}(E_{-\alpha_{i}})$ and degree $0$ to $H_{i}$ ($i=1,\dots, r$) so that the subset $\mathfrak{g}^{k}$ contains all the generators with degree $k$.

Let $\hat{\mathfrak g}$ be an affine Lie algebra associated with $\mathfrak{g}$.
The simple root system of $\hat{\mathfrak g}$ is obtained by adding the affine root $\alpha_{0}=-\theta$ to the simple roots of $\mathfrak{g}$,  where $\theta$ is the highest root of $\mathfrak{g}$. Then the Coxeter labels $n_{i}$ are defined to satisfy $\sum_{i=0}^{r}n_{i}\alpha_{i}=0$ and the Coxeter number is given by
\begin{equation}
    h=\sum_{i=0}^{r}n_{i}.
\end{equation}
The Coxeter numbers for the affine Lie algebras in this paper are listed in Appendix \ref{Appe.A}.

\subsection{The linear problems for affine Toda field equations}
We now discuss the linear problem associated with the affine Toda field equation. For an affine Lie algebra $\hat{\mathfrak{g}}$, one can consider two-dimensional affine Toda field theories \cite{Olive:1983mw, Wilson:1981zz, Leznov:1980ks, Mikhailov:1980my}, whose equation of motion is given by
\begin{equation}\label{eq: toda eq}
    \partial_{\Bar{z}}\partial_{z}\phi(z,\Bar{z})-\Big(\frac{m^{2}}{\beta}\Big)[\sum_{i=1}^{r}\alpha_{i}\exp\big(\beta\alpha_{i}\cdot \phi\big)+\alpha_{0}\exp\big(\beta\alpha_{0}\cdot \phi\big)]=0
\end{equation}
with $\phi(z,\Bar{z})=\sum_{i=1}^{r}\alpha_{i}^{\vee}\phi_{i}(z,\Bar{z})$.
Here $(z,\bar{z})$ are the complex coordinates of two-dimensional Euclidean spacetime, $\beta$ is a coupling constant, and $m$ is a mass parameter.
Eq. \eqref{eq: toda eq} can be written as the flatness condition $[{\cal L},\bar{\cal L}]=0$ for the 
Lax operators ${\cal L}$ and $\bar{\cal L}$, which are defined by
\begin{align}\label{eq:toda_lax}
\begin{split}
&\mathcal{L}=\partial_{z}+\beta\sum_{i=1}^{r}\partial_{z}\phi_{i}(z,\Bar{z})H_{i}+m\lambda\Lambda,\\
&\Bar{\mathcal{L}}=\partial_{\Bar{z}}+\lambda^{-1}e^{-\beta\sum_{i=1}^{r}\phi_{i}H_{i}}(m\Bar{\Lambda})\;e^{\beta\sum_{i=1}^{r}\phi_{i}H_{i}}.
\end{split}
\end{align}
Here $\lambda$ is a spectral parameter, $\Lambda=\sum_{i=0}^{r}E_{\alpha_{i}}$, and $\Bar{\Lambda}=\sum_{i=0}^{r}E_{-\alpha_{i}}$. 
The flatness condition implies the integrability condition of the linear problem
${\cal L}\Psi=\bar{\cal L}\Psi=0$.
For further convenience, we set the parameters  $\beta=m=1$.

One can take the conformal transformation \cite{Lukyanov:2010rn, Dorey:2012bx, Ito:2013aea, Adamopoulou:2014fca}.
\begin{equation}\label{eq: cft}
    z\rightarrow w(z),\quad \Bar{z}\rightarrow \Bar{w}(\Bar{z}),\quad \phi\rightarrow \hat{\phi}=\phi-\rho^{\vee}\log(\partial_{z}w\partial_{\Bar{z}}\Bar{w}),
\end{equation}
then Eq. \eqref{eq: toda eq} becomes so-called the modified affine Toda field equation
\begin{equation}\label{eq:matfes}
    \partial_{\Bar{z}}\partial_{z}\phi(z,\Bar{z})-[\sum_{i=1}^{r}\alpha_{i}\exp\big(\alpha_{i}\cdot \phi\big)+p(z)\Bar{p}(\Bar{z})\alpha_{0}\exp\big(\alpha_{0}\cdot \phi\big)]=0
\end{equation}
with
\begin{equation}
    p(z)=(\partial_{z}w)^{h},\quad \Bar{p}(\Bar{z})=(\partial_{\Bar{z}}\Bar{w})^{h}.
\end{equation}
Eq. \eqref{eq:matfes} can also be written as the flatness condition for
the Lax operators ${\cal L}_m$ and $\bar{\cal L}_m$, where
\begin{align}\label{eq: mdr lax}
\begin{split}
&\mathcal{L}_{m}=\partial_{z}+\sum_{i=1}^{r}\partial_{z}\phi_{i}(z)H_{i}+\lambda(\sum_{i=1}^{r}E_{\alpha_{i}}+p(z)E_{\alpha_{0}}),\\
&\Bar{\mathcal{L}}_{m}=\partial_{\Bar{z}}+\lambda^{-1}e^{-\sum_{i=1}^{r}\phi_{i}H_{i}}(\sum_{i=1}^{r}E_{-\alpha_{i}}+\Bar{p}(\Bar{z})E_{\alpha_{0}})e^{\sum_{i=1}^{r}\phi_{i}H_{i}}.
\end{split}
\end{align}
We can also consider the linear problem for modified Lax operators.
In particular, we will focus on its holomorphic part
\begin{equation}\label{eq:hol_lp1}
    \mathcal{L}_{m}\Psi(z,\Bar{z})=0.
\end{equation}
Let us analyze Eq. \eqref{eq:hol_lp1} explicitly for the affine Lie algebra $A_{r}^{(1)}$ \cite{Ito:2013aea, Adamopoulou:2014fca}. We choose the $(r+1)$-dimensional fundamental representation with the highest weight $\omega_1$, whose representation is given in Appendix \ref{Appe.A}. Then $\Psi(z,\Bar{z})=(\psi_1,\dots, \psi_{r+1})^{\text{T}}$ is a $(r+1)$-dimensional vector. The linear problem becomes
\begin{equation}\label{eq: Ar linear problem}
 \left\{ \partial_{z}+\begin{pmatrix}
        \partial_{z}\phi_{1} & \lambda &   &   &   \\
          & (\partial_{z}\phi_{2}-\partial_{z}\phi_{1}) & \lambda &   &   \\
          &   & \ddots &  &   \\
          &   &   &   (\partial_{z}\phi_{r}-\partial_{z}\phi_{r-1})     & \lambda \\
        \lambda p(z) &   &   &    &  -\partial_{z}\phi_{r}\\ 
    \end{pmatrix}
    \right\}
    \begin{pmatrix}
        \psi_{1}\\
        \psi_{2}\\
        \vdots\\
        \psi_{r}\\
        \psi_{r+1}\\
    \end{pmatrix}=0.
\end{equation}
After eliminating the components $\psi_{2},\dots,\psi_{r+1}$, we find the higher order ODE for $\psi_1(z,\bar{z})$
\begin{equation}\label{eq:full_Ar_ode}
    (-\lambda)^{-h}(\partial_{z}-\partial_{z}\phi_{r})(\partial_{z}+\partial_{z}\phi_{r}-\partial_{z}\phi_{r-1})\dots(\partial_{z}+\partial_{z}\phi_{2}-\partial_{z}\phi_{1})(\partial_{z}+\partial_{z}\phi_{1})\psi_{1}=p(z)\psi_{1}
\end{equation}
with $h=r+1$. In the same way, one can obtain the (pseudo) ordinary differential equations for classical twisted and untwisted affine Lie algebras \cite{Ito:2013aea}.
For $A_{2r-1}^{(2)}$, $B_{r}^{(1)}$, $D_{r+1}^{(2)}$, and $D_{r}^{(1)}$, they are 
\begin{align}\label{eq:other_ode}
\begin{split}
&A_{2r-1}^{(2)}:\; \lambda^{-(2r-1)}(\partial_{z}-\partial_{z}\phi_{1})\cdots (\partial_{z}-\partial_{z}\phi_{r}+\partial_{z}\phi_{r-1})(\partial_{z}+\partial_{z}\phi_{r}-\partial_{z}\phi_{r-1})\\
&\quad\quad\quad\quad \cdots (\partial_{z}+\partial_{z}\phi_{1})\psi+2\sqrt{p(z)}\partial_{z}\sqrt{p(z)}\psi=0,\\
&B_{r}^{(1)}:\; \lambda^{-2r}(\partial_{z}-\partial_{z}\phi_{1})\cdots (\partial_{z}-2\partial_{z}\phi_{r}+\partial_{z}\phi_{r-1})\partial_{z}(\partial_{z}+2\partial_{z}\phi_{r}-\partial_{z}\phi_{r-1})\\
&\quad\quad\quad\quad \cdots (\partial_{z}+\partial_{z}\phi_{1})\psi-4\sqrt{p(z)}\partial_{z}\sqrt{p(z)}\psi=0,\\
&D_{r+1}^{(2)}:\; \lambda^{-(2r+2)}(\partial_{z}-\partial_{z}\phi_{1})\cdots (\partial_{z}-2\partial_{z}\phi_{r}+\partial_{z}\phi_{r-1})\partial_{z}(\partial_{z}+2\partial_{z}\phi_{r}-\partial_{z}\phi_{r-1})\\
&\quad\quad\quad\quad \cdots (\partial_{z}+\partial_{z}\phi_{1})\psi-4p(z)\partial_{z}^{-1}p(z)\psi=0,\\
&D_{r}^{(1)}:\; \lambda^{-(2r-2)}(\partial_{z}-\partial_{z}\phi_{1})\cdots (\partial_{z}-\partial_{z}\phi_{r}-\partial_{z}\phi_{r-1}+\partial_{z}\phi_{r-2})\partial_{z}^{-1}\\
&\quad\quad\quad\quad (\partial_{z}+\partial_{z}\phi_{r}+\partial_{z}\phi_{r-1}-\partial_{z}\phi_{r-2})\cdots(\partial_{z}+\partial_{z}\phi_{1})\psi-4\sqrt{p(z)}\partial_{z}\sqrt{p(z)}\psi=0.\\
\end{split}
\end{align}
We can also construct the ODEs for $C_r^{(1)}$ and $A_{2r}^{(2)}$ but they are in the same form as $A_{r}^{(1)}$ type Eq. \eqref{eq:full_Ar_ode}.
In the next subsection, we will study the WKB analysis for the ODEs \eqref{eq:full_Ar_ode} and \eqref{eq:other_ode}.

If one further fixes the asymptotic behavior of $\phi_{i}$ at the origin $z=\bar{z}=0$ and takes the light-cone limit together with the conformal limit, Eq. \eqref{eq:full_Ar_ode} will become the one appearing in the ODE/IM correspondence \cite{Dorey:2006an}. To see this, set the functions $p(z)$ and $\Bar{p}(\Bar{z})$ to be
\begin{equation}
    p(z)=z^{hM}-s^{hM},\quad \Bar{p}(\Bar{z})=\Bar{z}^{hM}-\Bar{s}^{hM},
\end{equation}
for some positive real number $M>\frac{1}{h-1}$ and parameter $s$. For $z$, $\Bar{z}$ approaching $0$, we assume that the field $\phi_{i}$ is expanded as \cite{Lukyanov:2010rn}
\begin{gather}\label{eq: phi0}
  \phi_{i}(z,\bar{z}) = l_{i}\log(z\bar{z})+\mathcal{O}(1) , 
\end{gather}
where $l_{i}$ are constants. Now consider the light-cone limit $\Bar{z}\rightarrow0$ with $\phi_{i}$ in the form of Eq. \eqref{eq: phi0}. Then, we take the limit $z\sim s\rightarrow0$, $\lambda\rightarrow\infty$ to keep the combinations
\begin{equation}\label{eq: con_limit}
    x=\lambda^{\frac{1}{1+M}}z,\quad E=s^{hM}\lambda^{\frac{2M}{1+M}}
\end{equation}
finite. The ordinary differential equations of $\psi(x,E;\epsilon)$ now become
\begin{equation}
   (\partial_{x}-\frac{l_{r}}{x})(\partial_{x}+\frac{l_{r}}{x}-\frac{l_{r-1}}{x})\dots(\partial_{x}+\frac{l_{2}}{x}-\frac{l_{1}}{x})(\partial_{x}+\frac{l_{1}}{x})\psi=p(x,E)\psi
\end{equation}
with $p(x,E)=x^{hM}-E$, and the linear problem \eqref{eq: Ar linear problem} becomes
\begin{equation}
\mathcal{L}_{m}=\partial_{x}+\sum_{i=1}^{r}\frac{l_{i}}{x}+\sum_{i=1}^{r}E_{\alpha_{i}}+p(x,E)E_{\alpha_{0}}.
\end{equation}
Both the ODE and the linear problem can be used to show $A_{r}^{(1)}$-type ODE/IM correspondence. See \cite{Dorey:1999uk, Dorey:2007zx, Dorey:2006an} for more details.

\subsection{WKB analysis to the linear problems}\label{sec:wkb_an1}
We study the WKB analysis of the holomorphic linear problems \eqref{eq:hol_lp1} and the associated ordinary differential equation here. First, let us take the order parameter $\epsilon=\lambda^{-1}$ to follow the convention in the WKB method. We rescale the holomorphic modified Lax operator in Eq. \eqref{eq: mdr lax} as
\begin{equation}\label{eq:holo_mlax}    \epsilon\mathcal{L}_{m}=\epsilon\partial_{z}+\epsilon\sum_{i=1}^{r}\partial_{z}\phi_{i}(z,\Bar{z})H_{i}+\sum_{i=1}^{r}E_{\alpha_{i}}+p(z)E_{\alpha_{0}}.
\end{equation}
From now on, let us focus on the holomorphic coordinate $z$ and set $\bar{z}$ fixed. There are two approaches to the WKB analysis of the linear problem. Let us introduce both of them for $A_{r}^{(1)}$ type. 

The first approach is the so-called abelianization method \cite{Hollands:2019wbr}. It starts from the ordinary differential equation \eqref{eq:full_Ar_ode} with $\lambda=\epsilon^{-1}$:
\begin{equation}\label{eq:Ar_ode}
    [(-\epsilon)^{h}(\partial_{z}-\partial_{z}\phi_{r}(z))\cdots(\partial_{z}+\partial_{z}\phi_{2}(z)-\partial_{z}\phi_{1}(z))(\partial_{z}+\partial_{z}\phi_{1}(z))-p(z)]\psi_1(z,\epsilon)=0.
\end{equation}
To solve the ODE \eqref{eq:Ar_ode}, let us apply the WKB ansatz by setting 
\begin{equation}\label{eq: WKB expand}
    \psi_{1}(z,\epsilon)=\exp(\frac{1}{\epsilon}\int^z dz\,P(z,\epsilon))
\end{equation}
with $P(z,\epsilon)=\sum_{i=0}^{\infty}\epsilon^{i}P_{i}(z)$. The equation satisfied by $P(z,\epsilon)$ is called the Riccati equation. Let us call $P(z,\epsilon)$ the WKB solution for further convenience. One can calculate $P_{i}(z)$ recursively from the Riccati equation. 
The WKB analysis for $A_{r}^{(1)}$ type ODEs with $\phi_i=0$ have been done in \cite{Ito:2021boh} (see also \cite{Hollands:2019wbr,Yan:2020kkb}), where it has been shown there that $P_{i}(z)$ for odd $i$ and $i=2(r+1)k+r+2$ ($k=0,1,\dots$) are total derivatives. However, this is not true when $\phi_{i}\neq0$ as we will see in Section \ref{sect:Sec3}.

The second approach is applying the WKB method directly to the linear problem\footnote{For the ${\mathfrak sl}_3$ Miura opers related to affine Gaudin models, the WKB analysis has bee studied in \cite{Gaiotto:2020dhf}. }. For simplicity, let us set $\phi_{i}=0$. In the linear problem \eqref{eq: Ar linear problem}, we take the following WKB ansatz:
\begin{equation}\label{eq:wkb_ansatz1}
    \Psi(z,\epsilon)=\big(\exp(\frac{1}{\epsilon}\int^{z}P_{1}(z',\epsilon)dz'),\dots,\exp(\frac{1}{\epsilon}\int^{z} P_{2r+2}(z',\epsilon)dz')\big)^{\text{T}},
\end{equation}
where $P_{i}(z,\epsilon)=\sum_{n=0}^{\infty}\epsilon^{n}P_{i(n)}(z)$. 
In particular, $P_1(z,\epsilon)=P(z,\epsilon)$ with $\phi_{i}=0$.
After substituting the ansatz \eqref{eq:wkb_ansatz1} into the linear problem \eqref{eq:hol_lp1}, one find the following recursive relations
\begin{align}\label{eq: S recursion pertubation}
\begin{split}
    &P_{i+1(n)}=-(\frac{P_{i}'}{P_{i}})_{(n-1)}+P_{i(n)},\quad 1\leq i\leq r,\\
    &\sum_{i=1}^{r+1}(\frac{P_{i}'}{P_{i}})_{(n-1)}-\delta_{1n}(\frac{p'(z)}{p(z)})=0.
    \end{split}
\end{align}
which are equivalent to the Riccati equation for the ODE \eqref{eq:Ar_ode} with $\phi_{i}=0$ \cite{Ito:2021boh}
\begin{equation}
    (\epsilon\partial_{z}+P(z,\epsilon))^{r}P(z,\epsilon)+(-1)^{r}p(z)=0.
\end{equation}
This method is applicable to any affine Lie algebra, but rather cumbersome to find the solutions. Especially it can be used to solve the $D_{r+1}^{(2)}$-type pseudo-differential equations. See Appendix \ref{Appe.B} for more details.

So far, we have only considered $A_{r}^{(1)}$ types. The first WKB analysis approach can also be applied to $A_{2r-1}^{(2)}$, $B_{r}^{(1)}$-type ordinary differential equations in Eq. \eqref{eq:other_ode}. But it is still difficult to apply the same WKB analysis to $D_{r+1}^{(2)}$ and $D_{r}^{(1)}$-type differential equations for $\phi_i\neq0$, where the obtained ones include the pseudo-differential operators.
The second approach can be applied to linear problems with affine Lie algebras $A_{2r-1}^{(2)}$, $B_{r}^{(1)}$, and $D_{r+1}^{(2)}$ when $\phi_{i}=0$ but it is still challenging to solve the general $D_{r}^{(1)}$ type with $\phi_i\neq 0$. The WKB analysis of the linear problem in this section begins with the solution $\Psi(z,\epsilon)$. It inspires us that there may exist a method from the modified Lax operators \eqref{eq: mdr lax}. This is a central problem in the next section.

\section{Diagonalization of $A_{r}^{(1)}$-type linear  problem}\label{sect:Sec3}
In this section, we reconsider the WKB analysis of the linear problem from the viewpoint of the classical integrable models. The Lax formalism allows us to construct an infinite number of conserved charges algebraically. For the Lax equations associated with affine Lie algebras, Drinfeld and Sokolov have diagonalized the Lax operator in terms of the eigenstates of $\Lambda$ in Eq. \eqref{eq:toda_lax} and systematically constructed the conserved charges \cite{Drinfeld:1984qv}. In the diagonalized basis, the linear problem reduces to the set of independent first-order differential equations which can be solved easily.

\subsection{Integrable hierarchies and diagonalized linear problems}
We first illustrate a diagonalization procedure for the $A_r^{(1)}$-type linear problem. Let us consider the linear problem ${\cal L}\Psi=0$ in the $(r+1)$-dimensional representation. Here ${\cal L}$ is a general form of Eq. \eqref{eq:toda_lax}
\begin{equation}\label{eq:DS_lax}
    \mathcal{L}=\partial_{z}+q(z,\lambda)+\lambda\Lambda,
\end{equation}
where $q(z,\lambda)$ is a $(r+1)\times (r+1)$ lower triangular matrix, and $\Lambda=\sum_{i=0}^{r}E_{\alpha_i}$ which is invertible.  
If  $q(z)$ depends on a time coordinate $t$ and satisfies the Lax equation $\frac{d{\cal L}}{dt}=[\mathcal{A}, {\cal L}]$ for a specific differential operator $\mathcal{A}$, one can define a classical integrable system with infinite conserved charges. 
There exists a formal series $T$ such that the operator $\mathcal{L}_{0}=T\mathcal{L}T^{-1}$ has the form \cite{Drinfeld:1984qv}
\begin{equation}\label{eq: DS dlax}
    \mathcal{L}_{0}=\partial_{z}+\lambda\Lambda+\sum_{i=0}^{\infty}\lambda^{-i}I_{i}(z)\Lambda^{-i}
\end{equation}
with
\begin{equation}
    T(z,\lambda)=E+\sum_{i=1}^{\infty}\lambda^{-i}h_{i}(z)\Lambda^{-i},
\end{equation}
where $E$ is the identity matrix, $I_{i}(z)$ are functions of $z$, and $h_{i}(z)$ are diagonal matrices whose entries are functions of $z$. The series $T(z,\lambda)$ is determined up to multiplication on the left by series of the form $d(z)=E+\sum_{i=1}^{\infty}\lambda^{-i}d_{i}(z)\Lambda^{-i}$, where $d_{i}(z)$ are functions of $z$. The form of Eq. \eqref{eq: DS dlax} implies that ${\cal L}_0$ is a diagonalized Lax operator for the eigenvectors of $\Lambda$. $I_i(z)$ are argued to be the conserved charge densities of the integrable model, which will be discussed in Section \ref{Sec_5}. 
For the $(r+1)$-dimensional fundamental representation of $A_r$, the eigenvalues of $\Lambda$ are $1,\omega,\dots, \omega^{r}$ with $\omega=\exp({2\pi i\over r+1})$.
We can obtain $I_{i}(z)$ and $h_{i}(z)$  recursively by expanding $\mathcal{L}_{0}T=T\mathcal{L}$ in $\lambda^{-1}$ and comparing the coefficients of $\lambda^{-i}$. For $i=0$ and $i\geq 1$, we find
\begin{align}\label{eq: DS rec}
    \begin{split}
        &\quad\quad\quad\quad\quad\quad\quad\quad\quad\quad h_{1}^{\sigma}(z)+I_{0}(z)E=q_{0}(z),\\
        &h_{i+1}(z)-h_{i+1}^{\sigma}(z)-I_{i}(z)E=\partial_{z}h_{i}(z)+\sum_{j=0}^{i-1}\big(I_{j}(z)E-q_{j}(z)\big)h_{i-j}(z)-q_{i}(z),
    \end{split}
\end{align}
where $q(z)=\sum_{i=0}^{\infty}\lambda^{-i}q_{i}(z)\Lambda^{-i}$ with $q_{i}(z)$ diagonal matrices, and $h_{i}^{\sigma}=\Lambda h_{i}\Lambda^{-1}$. $I_{i}$ is uniquely determined up to total derivatives due to the choice of $d(z)$ in $T(z,\lambda)$. From the relation $\mathcal{L}_{0}=T\mathcal{L}T^{-1}$, we obtain
\begin{equation}\label{eq: traceless condition}
    \text{Tr}(\lambda^{-i}\sum_{i=0}^{\infty}(I_{i}(z)-q_{i}(z))\Lambda^{-i})=-\text{Tr}(\partial_{z}TT^{-1})=-\partial_{z}\ln\det T,
\end{equation}
which changes up to $-\partial_{z}\ln\det d(z)$ with further multiplication on the left of $T$ by $d(z)$.
After the diagonalization of $\Lambda$, Eq. \eqref{eq: DS dlax} can be written as
\begin{align}\label{eq:DS_dilax}
    \begin{split}
\mathcal{L}_{\text{diag}}&=\partial_{z}+\lambda\Lambda_{\text{diag}}+\sum_{i=0}^{\infty}\lambda^{-i}I_{i}(z)\Lambda_{\text{diag}}^{-i}\\
        &=\partial_{z}+\lambda\Lambda_{\text{diag}}+
        \begin{pmatrix}
            I(z,e^{\frac{2\pi i }{r+1}}\lambda) &  &  &  & \\
			& \ddots &  &  & \\
			& & I(z,e^{\frac{2\pi i (r-1)}{r+1}}\lambda) &   & \\
			&  &  & I(z,e^{\frac{2\pi i r}{r+1}}\lambda) &   \\
			&  &  &  &  I(z,\lambda)\\
        \end{pmatrix},
    \end{split}
\end{align}
where $\Lambda_{\text{diag}}=\textbf{Diag}\{e^{\frac{2\pi i r}{r+1}}, e^{\frac{2\pi i(r-1) }{r+1}},e^{\frac{2\pi i(r-2) }{r+1}},\dots,1\}$ and $I(z,\lambda)=\sum_{i=0}^{\infty}I_{i}(z)\lambda^{-i}$.  The order of eigenvalues in $\Lambda_{\text{diag}}$ and $I(z,e^{\frac{2\pi i n }{r+1}}\lambda)$ in Eq. \eqref{eq:DS_dilax} is not fixed.

The gauge transformation and diagonalization of the Lax operator \eqref{eq:DS_lax} based on $A_r^{(1)}$ can be generalized to an affine Lie algebra $\hat{\mathfrak g}$ and its arbitrary representation.
In general, $q(z)$ is decomposed as $q(z) \in \oplus_{i=\infty}^{0}{\mathfrak g}^{-i}$ and $\Lambda=\sum_{i=0}^{r}E_{\alpha_i}$  admits zero eigenvalues. 
After the gauge transformation by $T$, the Lax operator ${\cal L}$ becomes
\begin{equation}\label{eq: DS dlax gr}
    \mathcal{L}_{0}=T\mathcal{L}T^{-1}=\partial_{z}+\lambda\Lambda+H(z,\lambda)
\end{equation}
with
\begin{equation}
    T(z,\lambda)=e^{U(z,\lambda)},\quad U(z,\lambda)=\sum_{i=1}^{\infty}\lambda^{-i}U^{i}(z),\quad U^i(z)\in {\mathfrak g}^{-i}
\end{equation}
and $H(z,\lambda)\in \text{Ker ad}\Lambda$ satisfies $[H,\Lambda]=0$. For the fundamental representation with the highest weight $\omega_1$ for $\hat{\mathfrak g}=B_{r}^{(1)},D_{r+1}^{(2)},A_{2r-1}^{(2)}$, and $D_{r}^{(1)}$, $H(z,\lambda)$ is given by \cite{Drinfeld:1984qv}
\begin{align}\label{eq:h1}
    &H(z,\lambda) = \sum_{i=0}^{\infty}\lambda^{-(2i+1)}I_{i}(z)\Lambda^{-(2i+1)}, \quad\text{for}\; B_{r}^{(1)},\; D_{r+1}^{(2)},\;A_{2r-1}^{(2)},
\end{align}
\begin{align}\label{eq:h2}
    &H(z,\lambda) =\sum_{i=0}^{\infty}\lambda^{-(2i+1)}I_{i}(z)\Lambda^{-(2i+1)}+\sum_{i=0}^{\infty}\lambda^{-(2i+1)}J_{i}(z)F,  \quad\text{for}\; D_{r}^{(1)}. 
\end{align}
Due to the existence of zero eigenvalues in $\Lambda$, we define $\Lambda^{-(2i+1)}=\Lambda^{-(2i+1)+kh}$ with sufficiently large $k$ for $B_{r}^{(1)}$, $A_{2r-1}^{(2)}$, and $D_{r}^{(1)}$. In Eq. \eqref{eq:h2}, there is an extra matrix $F$ commuting with $\Lambda$, which generates a new set of conserved densities in $D_{r}^{(1)}$ case corresponding to two zero eigenvalues in $\Lambda$ \cite{He:2009dfe}.
$I_i$, $J_i$, and $U^i$ can be determined recursively. However, $U(z,\lambda)$ is not fixed uniquely because the form of $\mathcal{L}_{0}$ is preserved after the further gauge transformation $e^{\textbf{ad}U'}$ ($U'\in \text{Ker ad}\Lambda$). 
Up to total derivatives, $I_i$ and $J_i$ are independent of different choices of $U(z)$. 
The terms proportional to $\Lambda^{-2i}$ do not appear in Eq. \eqref{eq:h1} and Eq. \eqref{eq:h2} since their coefficients are total derivatives. 
Eq. \eqref{eq: DS dlax gr} can be understood as diagonalization of ${\cal L}$ in terms of eigenvectors of $\Lambda$.

\subsection{The diagonalization approach to the linear problem}\label{SSEC: tda}
We have seen that the Lax operator \eqref{eq:DS_lax} is diagonalized after the gauge transformation, and the linear problem in the diagonalized form can be solved easily. Here we apply this approach to the linear problem ${\cal L}_m\Psi=0$ in Eq. \eqref{eq:holo_mlax} associated with the modified affine Toda field equation, which provides a way to find the WKB expansion to the solution. In the previous subsection, we have obtained diagonalized Lax operators for general affine Lie algebras. It is interesting to apply the diagonalization method to the linear problem for the modified affine Toda field equation.

Although the approach based on the recursive relation like Eq. \eqref{eq: DS rec} is enough to perform the diagonalization, it is rather complicated to find the concrete form of $T(z)$. We will investigate a direct approach to diagonalize the linear problem. Let us introduce the parameter $\epsilon$ and begin with the modified Lax operator in Eq. \eqref{eq:holo_mlax}
\begin{equation}    \epsilon\mathcal{L}_{m}=\epsilon\partial_{z}+\epsilon\sum_{i=1}^{r}\partial_{z}\phi_{i}(z)H_{i}+\sum_{i=1}^{r}E_{\alpha_{i}}+p(z)E_{\alpha_{0}}.
\end{equation}
One can view the Lax operator as a covariant derivative with corresponding connection $A(z)=\sum_{i=1}^{r}\epsilon\partial_{z}\phi_{i}(z)H_{i}+\sum_{i=1}^{r}E_{\alpha_{i}}+p(z)E_{\alpha_{0}}$. Then the gauge transformation is given by
\begin{equation}
\textbf{Gau}_{T}[A(z)]=T^{-1}(z)A(z)T(z)+\epsilon T^{-1}(z) \partial_{z} T(z).
\end{equation}
The next step is to find a matrix $T(z)$ to perform the diagonalization. In this paper, we diagonalize $A(z)$ row by row from bottom to top, which is regarded as a generalization of Section 11.1 in \cite{Babelon:2003qtg}. The matrix is given by
\begin{equation}\label{transfer Mat}
    T(z)=T_{d}T_{d-1}\dots T_{3}T_{2}T_{1},
\end{equation}
where $d$ is the dimension of the representation and $T_{i}(z)$ are $d\times d$ matrices satisfying
\[   
T_{i}(z)_{ab} = 
     \begin{cases}
       1, &\quad\text{if}\quad a=b,\\
       g_{i,b}(z),  &\quad\text{if}\quad a=i,\quad b\neq i,\quad 1\leq b\leq d,\\
       0, &\quad\text{otherwise.}\\
     \end{cases}
\]
For each step of the gauge transformation $\textbf{Gau}_{T_{i}}$, we fix $g_{i,b}(z)$ such that the connection $A'(z)$ satisfies
\begin{equation}\label{diag eq}
    A'_{ij}=0,\quad 1\leq j\leq d,\quad j\neq i.
\end{equation}
There are finally $d-1$ constraints in Eq. \eqref{diag eq} to diagonalize the $i$-th row in $A(z)$ and fix the diagonal elements perturbatively. Except for the $D_{r}^{(1)}$ type, the $(d-1)$ constraints in a row are reduced to a single Riccati equation\footnote{For $D_{r}^{(1)}$ type, the constraints are reduced to two Riccati equations. We shall discuss the details in Section \ref{sect:Sec_4}.}. The final diagonalized connection $ A_{\text{diag}}(z)$ is given by
\begin{equation}\label{eq: diag appro}
    A_{\text{diag}}(z)=\textbf{Gau}_{T_{1}}\circ\textbf{Gau}_{T_{2}}\dots\textbf{Gau}_{T_{d-2}}\circ\textbf{Gau}_{T_{d-1}}\circ\textbf{Gau}_{T_{d}}[A(z)].
\end{equation}
Due to the traceless condition, $d-1$ Riccati equations are necessary to determine all the diagonal elements. The diagonal elements are uniquely determined up to total derivatives. One can further act $T_{\textbf{diag}}=\textbf{diag}\{\exp(t_{1}(z,\epsilon)),\dots,\exp(t_{d}(x,\epsilon))\}$ after $T(z)$, then the diagonal element $[A_{\textbf{diag}}(z)]_{ii}$ will shift by $\partial_{z}t_{i}(z)$.
In the following, we will explain the diagonalization procedure for lower-rank examples.

\subsection{The diagonalization of $A_{1}^{(1)}$}\label{sec_3.3}
The $A_{1}^{(1)}$ affine Toda field theory is also called the Sinh-Gordon model. Its diagonalization has been studied in \cite{Babelon:2003qtg}. The approach there can be viewed as a particular case of ours. The holomorphic part of the modified Lax operator \eqref{eq: mdr lax} is of the form
\begin{equation}
    \mathcal{L}_{m}=\epsilon\partial_{z}+ \epsilon\partial_{z}\phi(z)H+E_{\alpha}+p(z)E_{-\alpha}
\end{equation}
with
\begin{equation}
    H=\left(
\begin{array}{cc}
 1 & 0 \\
 0 & -1\\
\end{array}
\right),\quad
E_{\alpha}=\left(
\begin{array}{cc}
 0 & 1 \\
 0 & 0\\
\end{array}
\right),\quad
E_{-\alpha}=\left(
\begin{array}{cc}
 0 & 0 \\
 1 & 0\\
\end{array}
\right).
\end{equation}
We define a matrix: $T(z)=T_{1}T_{2}$ of the gauge transformation $\mathcal{L}\rightarrow \mathcal{L}'\equiv T^{-1}\mathcal{L}T$ 
by
\begin{equation}
    T_{2}(z)=\left(\begin{array}{cc}
 1 & 0 \\
 g_{2,1}(z,\epsilon) & 1 \\
\end{array}
\right),\quad
T_{1}(z)=\left(\begin{array}{cc}
 1 & g_{1,2}(z,\epsilon) \\
 0 & 1\\
\end{array}
\right).
\end{equation}
$T_{2}(z)$ is determined to diagonalize the second row
\begin{equation}
    {\bf Gau}_{T_2}[A(z)]=\left(
\begin{array}{cc}
 g_{2,1}+\phi'  & 1 \\
 -2 \epsilon  g_{2,1} \epsilon\phi '+\epsilon  g_{2,1}'-g_{2,1}^2+p & -g_{2,1}-\epsilon\phi' \\
\end{array}
\right).
\end{equation}
It gives the  condition for $g_{2,1}(z)$:
\begin{equation}\label{eq: A11 Ricatti}
    g_{2,1}^2(z,\epsilon)+2\epsilon g_{2,1}(z,\epsilon)\phi'(z,\Bar{z})-\epsilon  g_{2,1}'(z,\epsilon)-p(z)=0.
\end{equation}
The diagonal element $f(z,\epsilon):=-g_{2,1}(z,\epsilon)-\epsilon \phi'(z)$ satisfies
\begin{equation}\label{eq: A11 mono Ricatti}
    f^2(z,\epsilon)+\epsilon f'(z,\epsilon)-\epsilon ^2 u_{2}(z)-p(z)=0
\end{equation}
with $u_{2}(z)=\phi'(z)^2-\phi''(z)$. After the diagonalization of the second row, 
the second gauge transformation $T_{1}(z)$ gives
\begin{equation}
    {\bf Gau}_{T_1}\circ {\bf Gau}_{T_2}[A(z)]=\left(
\begin{array}{cc}
 -f(z,\epsilon) & 1-2g_{1,2}(z,\epsilon)f(z,\epsilon) \\
 0 & f(z,\epsilon) \\
\end{array}
\right).
\end{equation}
We do not need to extract the diagonalization condition from the first row since $g_{1,2}$ is independent of the diagonal elements. Let us substitute $f(z,\epsilon)=\sum_{i=0}^{\infty}\epsilon^{i}f_{i}(z)$ into Eq. \eqref{eq: A11 Ricatti}, then Eq. \eqref{eq: A11 mono Ricatti} can be solved perturbatively. The first four orders of diagonal elements $f(z,\epsilon)$ are listed below
\begin{align*}
        f_{0}(z)&=\sqrt{p(z)},\\
        f_{1}(z)&=-\frac{1}{2}\partial_{z}\ln f_{0},\\
        f_{2}(z)&=\frac{f_0''}{16 f_{0}^{2}}+\frac{u_{2}(z)}{2f_{0}}+\partial_{z}(\frac{3 f'_{0}}{16 f_{0}^{2}}),\\
        f_{3}(z)&=\partial_{z}(\frac{u_{2}(z)}{4f_{0}^{2}}-\frac{3 f_{0}^{'2}}{16 f_{0}^{4}}+\frac{f^{''}_{0}}{8f_{0}^{3}}).
\end{align*}
Besides, the diagonal elements are uniquely determined up to total derivative terms like Eq. \eqref{eq: traceless condition}. Therefore, the diagonal elements can be summarized as
\begin{equation}\label{eq: A1 dlax}
    A_{\text{diag}}(z)=\left(
\begin{array}{cc}
 -f(z,-\epsilon)+d(*) & 0\\
 0 & f(z,\epsilon)\\
\end{array}
\right),
\end{equation}
where $d(*)$ denotes total derivatives, and the form of the first diagonal element can be seen from 
\begin{align*}
       -f_{0}(z)&=-\sqrt{p(z)},\\
        -f_{1}(z)&=f_{1}(z)+\partial_{z}\ln f_{0},\\
        -f_{2}(z)&=-f_{2}(z),\\
        -f_{3}(z)&=f_{3}(z)-\partial_{z}(\frac{u_{2}(z)}{2f_{0}^{2}}-\frac{3 f_{0}^{'2}}{8 f_{0}^{4}}+\frac{f^{''}_{0}}{4f_{0}^{3}}).
\end{align*}

Next, we show the relations between $f(z,\epsilon)$ and the WKB solutions $P(z,\epsilon)$ in $A_{r}$-type ordinary differential equations \eqref{eq:Ar_ode}. The $A_{1}^{(1)}$-type ordinary differential equation is given by
\begin{equation}\label{eq: A11 ode}
    \big(\epsilon^{2}\partial_{z}^{2}+\epsilon^{2}\partial_{z}^{2}\phi(z)-\epsilon^{2}(\partial_{z} \phi)^{2}-p(z)\big)\psi(z,\epsilon)=0.
\end{equation}
Substituting the WKB ansatz \eqref{eq: WKB expand}, one can obtain the Riccati equation for $P(z,\epsilon)$
\begin{equation}\label{eq: A11 riccati}
    P^2(z,\epsilon)+\epsilon P'(z,\epsilon)-\epsilon ^2 \big(\phi'(z)^2-\phi''(z)\big)-p(z)=0.
\end{equation}
The perturbative solution can be given by substituting the expansion $P(z,\epsilon)=\sum_{i=0}^{\infty}\epsilon^{i}P_{i}(z)$ to Eq. \eqref{eq: A11 riccati}. The Riccati equation here is the same as Eq. \eqref{eq: A11 mono Ricatti}. After setting $P_{0}(z)=\sqrt{p(z)}$, one can find the equality between $f_{i}(z)$ and $P_{i}(z)$. 

Finally, let us also compare $f(z,\epsilon)$ with $I(z,\lambda)$ in  Eq. \eqref{eq: DS dlax}. Due to the existence of $\epsilon$ in front of $\partial_{z}$ in Eq. \eqref{eq:holo_mlax}, $f_{0}(z)$ is actually the (-1)-th term corresponding to the $\lambda\Lambda_{\text{diag}}$ term. Besides this, $f_{i}(z)$ here corresponds to $I_{i-1}(z)$ in Eq. \eqref{eq: DS dlax}. Especially, if one takes the inverse conformal transformation, namely $p(z)=1$, $S_{0}(z)=1$ matches the result $\lambda\Lambda_{\text{diag}}$. Actually, $f(z,\epsilon)$ and $I(z,\lambda)$ turn out to be the same set of conserved densities, which will be the main part of Section \ref{Sec_5}.

\subsection{The diagonalization of $A_{2}^{(1)}$}\label{sec_3.4}
The equality $f_{i}(z)=P_{i}(z)$ can be generalized into general $A_{r}^{(1)}$ with $r\geq 2$. As a trial, let us further investigate the $A_{2}^{(1)}$ case. The diagonalization approach in Eq. \eqref{SSEC: tda} can be applied directly. The modified Lax operator is
\begin{equation}
\epsilon\mathcal{L}_{m}=\epsilon\partial_{z}+\sum_{i=1}^{2}\epsilon\partial_{z}\phi_{i}(z,\Bar{z})H_{i}+\sum_{i=1}^{2}E_{\alpha_{i}}+p(z)E_{\alpha_{0}}
\end{equation}
with the representation of $A_{2}^{(1)}$ in Appendix \ref{Appe.B}. 

We perform the gauge transformation with $T=T_3 T_2 T_1$, from which we find $A_{\text{diag}}(z)$. The first gauge transformation by $T_{3}$ leads to
\begin{align}
    {\bf Gau}_{T_3}[A(z)]&=
    \begin{pmatrix}
    \epsilon \phi'_1 & 1 & 0\\
    g_{3,1} & g_{3,2}+\epsilon (\phi'_2-\phi'_1) & 1\\
    {\bf Gau}_{T_3}[A(z)]_{3,1} & {\bf Gau}_{T_3}[A(z)]_{3,2} & -g_{3,2} -\epsilon\phi'_2
    \end{pmatrix},
\end{align}
where
\begin{align}
\begin{split}
    &{\bf Gau}_{T_3}[A(z)]_{3,1}=-g_{3,1} \left(g_{3,2}+\epsilon  \left(\phi _1'+\phi _2'\right)\right)+\epsilon  g_{3,1}'+p,\\
    &{\bf Gau}_{T_3}[A(z)]_{3,2}=\epsilon  g_{3,2} \left(\phi _1'-2 \phi _2'\right)+\epsilon  g_{3,2}'-g_{3,2}^2-g_{3,1}. %,\\
    %&{\bf Gau}_{T_3}[A(z)]_{3,3}=-g_{1,2}-\epsilon  \phi _2'.
\end{split}
\end{align}
Let us set the diagonal element $f(z,\epsilon)\equiv -g_{3,2}-\epsilon  \phi'_2$. The diagonalization condition ${\bf Gau}_{T_3}[A(z)]_{3,1}={\bf Gau}_{T_3}[A(z)]_{3,2}=0$ solves $g_{3,1}$ and gives the equation for $g_{3,2}$ or $f(z,\epsilon)$. 

After the second gauge transformation $T_{2}$, the connection becomes
\begin{align}
    {\bf Gau}_{T_2T_3}[A(z)]&=
    \begin{pmatrix}
    \epsilon \phi'_1 +g_{2,1}& 1 & g_{2,3}\\
    {\bf Gau}_{T_2T_3}[A(z)]_{2,1} & g_{3,2}+\epsilon(\phi'_2-\phi'_1)-g_{2,1} & {\bf Gau}_{T_2T_3}[A(z)]_{2,3}\\
    0 & 0 & f
    \end{pmatrix},
\end{align}
where
\begin{align}
    \begin{split}
        &{\bf Gau}_{T_{2}T_{3}}[A(z)]_{2,1}=-2 \epsilon  g_{2,1} \phi _1'+\epsilon  g_{2,1} \phi _2'+\epsilon  g_{2,1}'-g_{2,1}^2+g_{3,2} g_{2,1}+g_{3,1},\\
        &{\bf Gau}_{T_{2}T_{3}}[A(z)]_{2,3}=-\epsilon  g_{2,3} \phi _1'+2 \epsilon  g_{2,3} \phi _2'+\epsilon  g_{2,3}'+2 g_{3,2} g_{2,3}-g_{2,1} g_{2,3}+1.\\
    \end{split}
\end{align}
We impose the conditions ${\bf Gau}_{T_{2}T_{3}}[A(z)]_{2,1}={\bf Gau}_{T_{2}T_{3}}[A(z)]_{2,3}=0$ and set $h(z,\epsilon)\equiv g_{2,1}+\epsilon  \phi _1'$. Eq. ${\bf Gau}_{T_{2}T_{3}}[A(z)]_{2,1}=0$ leads to the equation for $h(z,\epsilon)$, while
$g_{2,3}$ is determined by solving ${\bf Gau}_{T_{2}T_{3}}[A(z)]_{2,3}=0$.

Gauge transformation by $T_1$ eliminates the off-diagonal elements in the first row. Finally, we obtain the diagonalized connection $A_{\text{diag}}(z)={\bf Gau}_{T_1T_{2}T_{3}}[A(z)]$.
$A_{\text{Diag}}(z)$ becomes of the form $\textbf{diag}\{h(z,\epsilon),-f(z,\epsilon)-h(z,\epsilon),f(z,\epsilon)\}$ satisfy the Riccati equations:
\begin{align}\label{eq:A21_Ric}
\begin{split}
    &f^3+3 \epsilon  f f'-\epsilon ^{2}u_{2}f+\epsilon ^{2}f'' -\epsilon^3 u_{3}-p=0,\\
    &h^2+fh+f^2-\epsilon  h'+\epsilon  f'-\epsilon ^2 u_2=0.\\
    \end{split}
\end{align}
with
\begin{align}\label{eq:A2miura}
    \begin{split}
        &u_{2}(z)=\phi_1'(z)^{2}-\phi_2'(z)\phi_1'(z)+\phi_2'(z)^{2}-\phi_1''(z)-\phi_2''(z),\\
        &u_{3}(z)=2 \phi_2'(z) \phi_2''(z)-\phi_1'(z) \phi_2''(z)-\phi_1'(z) \phi_2'(z)^2+\phi_1'(z)^2 \phi_2'(z)-\phi_{2}^{(3)}(z).
    \end{split}
\end{align}
The second equation implies that $h(z,\epsilon)$ depends on $f(z,\epsilon)$, while the first  equation can be obtained from that of $P(z,\epsilon)$ for the $A_{2}^{(1)}$ ordinary differential equation of $\psi(z,\epsilon)=\exp(\frac{1}{\epsilon}\int dz\,P(z,\epsilon))$
\begin{equation}\label{eq:A21_ode}
    (-\epsilon)^{3}(\partial_{z}-\partial_{z}\phi_{1})(\partial_{z}-\partial_{z}\phi_{2}+\partial_{z}\phi_{1})(\partial_{z}+\partial_{z}\phi_{2})\psi+p(z)\psi=0.
\end{equation}
Eq. \eqref{eq:A21_ode} is the adjoint ODE of Eq. \eqref{eq:Ar_ode} with $r=2$, where the adjoint means $\partial_{z}\rightarrow-\partial_{z}$ and $\phi_{i}\rightarrow\phi_{h-i}$.
We can solve the coupled Riccati equations \eqref{eq:A21_Ric} perturbatively. Let us expand $f$ and $h$ as $f=\sum_{n=0}^{\infty}f_n \epsilon^n$ and $h=\sum_{n=0}^{\infty}h_n \epsilon^n$. The leading terms are $f_0$ and $h_0$, which can be obtained by solving $f_0^3=p$ and $h_0^2+f_0h_0+f_0^2=0$. The result is
\begin{align}
f_0&= p^{\frac{1}{3}},\quad h_0= e^{\pm {2\pi i\over3}}f_0.
\end{align}
Here, we take the minus sign in the second equation then the higher-order terms are represented in terms of $f_{0}$:
\begin{align*}
        f_{1}(z)&=-\frac{f_{0}'}{f_{0}}, & h_{1}(z)&=f_1(z) %\frac{f_{0}'}{f_{0}}
        +2\partial_{z}(\ln f_{0}),\\
        f_{2}(z)&=\frac{f_{0}''}{6f_{0}^{2}}+\frac{u_{2}(z)}{3f_{0}}+\partial_{z} (\frac{ f'_{0}}{2f_{0}^{2}}), & h_{2}(z)&=e^{{2\pi i\over3}}f_{2}(z),\\
        f_{3}(z)&=-\frac{u_{3}(z)}{3f_{0}^{2}}+\frac{f'_{0}u_{2}(z)}{3f_{0}^{3}}-\partial_{z}(-\frac{f_{0}^{'2}}{2 f_{0}^{4}}+\frac{f^{''}_{0}}{3f_{0}^{3}}-\frac{u_{2}(z)}{3S_{0}^{2}}), & h_{3}(z)&=e^{{4\pi i\over3}}(f_{3}(z)-\partial_{z}(\frac{u_{2}}{3f_{0}^{2}})).
\end{align*}
It is shown $f_{i}$ and $h_{i}$, up to total derivatives, are connected by phase relations $e^{{2\pi in\over3}}\,(n=0,1,2)$ which is expected from the structure in \eqref{eq: DS dlax}. The diagonal connection is summarized as
\begin{equation}\label{ma A21}
    A_{\text{diag}}(z)=\begin{pmatrix}
 e^{-\frac{i2\pi}{3}}f(z,e^{\frac{i2\pi}{3}}\epsilon) +d(*)& 0 & 0 \\
 0 & e^{-\frac{i4\pi}{3}}f(z,e^{\frac{i4\pi}{3}}\epsilon) +d(*) & 0 \\
 0 & 0 & f(z,\epsilon)  \\
\end{pmatrix},%+\text{total derivatives}.
\end{equation}
where $d(*)$ denotes total derivatives. The order of the three diagonal elements is not fixed but determined by their lowest order in $\epsilon$. For instance, the first two diagonal elements will be exchanged if one takes $h_0= e^{{2\pi i\over3}}f_0$ instead of $h_0= e^{-{2\pi i\over3}}f_0$.

After solving the three independent linear equations $(\epsilon \partial +A_{\text{diag}})\Psi'=0$, one obtains the solution of the linear problem by the gauge transformation $\Psi=T\Psi'$.

Finally, there is one comment on $u_{i}(z)$. The relations between $u_{i}(z)$ and  $\partial_{z}\phi_{i}(z)$ are obtained by the Miura transformation. There exists another equivalent canonical Lax operator in terms of $u_{i}(z)$:
\begin{equation}\label{eq: mcan Lax}
\epsilon\mathcal{L}_{\text{can}}=\epsilon\partial_{z}+\sum_{i=1}^{r}\epsilon^{i+1}u_{1+i}(z)e_{i+1,1}+\sum_{i=1}^{r}E_{\alpha_{i}}+p(z)E_{\alpha_{0}}
\end{equation}
with $(e_{i,j})_{a,b}=\delta_{i,a}\delta_{j,b}$. One can obtain the same Riccati equation after the same diagonalization method. In Section \ref{Sec_5}, we will discuss the canonical Lax operator in detail and utilize it to derive the WKB solutions from conserved densities. Before that, let us generalize the diagonalization to other affine Toda field theory types in the following subsection and Section \ref{sect:Sec_4}.

\subsection{Generalization to $A_{r}^{(1)}$}
In the last two subsections, we only show the diagonal results of $A_{1}^{(1)}$ and $A_{2}^{(1)}$, but the method can be applied to general $A_{r}^{(1)}$ directly. So far, we have checked the results until $A_{5}^{(1)}$ case. Let us consider the linear problem for $A_r^{(1)}$ in the fundamental representation with dimension $h=r+1$. After the step-by-step diagonalization procedure for the connection $A(z)$, the functions $g_{i,j}$ $(i>j)$ in the gauge transformations obey a set of non-linear equations, which reduce to the Riccati equations of the diagonal components in $A_{\text{diag}}$. The bottom component of the diagonal element $A_{\text{diag}}(z)$ obeys the Riccati equation of the ordinary differential equations 
\begin{equation}\label{eq:ad_Ar_ode}
    (-\epsilon)^{h}(\partial_{z}-\partial_{z}\phi_{1})(\partial_{z}-\partial_{z}\phi_{2}+\partial_{z}\phi_{1})\cdots(\partial_{z}+\partial_{z}\phi_{r})\psi(z,\epsilon)=p(z)\psi(z,\epsilon),
\end{equation}
which is adjoint of the ODE \eqref{eq:full_Ar_ode}. One can solve the set of Riccati equations perturbatively as we did in the $A_2^{(1)}$ case. It turns out that other diagonal elements in the connection can be summarized as the phase rotation
of $\epsilon$ with one of the eigenvalues of $\Lambda_{0}$: $\{1, e^{\frac{2\pi i }{h}},e^{\frac{4\pi i }{h}},\dots,e^{\frac{2\pi i r}{h}}  \}$ up to total derivatives. Then the diagonalized connection $A_{\textbf{diag}}$ can be of the form
\begin{align}\label{eq:Ar_Adiag}
\textbf{Diag}\{e^{-{2\pi i\over h}}f(z,e^{{2\pi i\over h}}\epsilon)+d(*),\dots, e^{-{2\pi ir\over h}}f(z,e^{{2\pi ir\over h}}\epsilon)+d(*),f(z,\epsilon)\}.
\end{align}
As we have mentioned in $A_{2}^{(1)}$ type, the phase in front of a diagonal element is still not fixed but determined by the lowest order in $\epsilon$. 

The traceless condition does not change up to total derivatives after the gauge transformation:
\begin{equation}
    \sum_{n=0}^{h-1}e^{-\frac{2\pi i n}{h}}f(x,e^{\frac{2\pi i n}{h}}\epsilon)=d(*),
\end{equation}
which implies that the $(1+hk)$-th term ($k\in\mathbf{Z}$) should be the total derivative and it agrees with the results in \cite{Ito:2021boh}.

\section{Diagonalization of other linear problems}\label{sect:Sec_4}
Since our diagonalization approach is quite general, we can apply it to the linear problems associated with other classical affine Lie algebras and study their WKB solutions. Especially, the algebras $D_{r}^{(1)}$ and $D_{r+1}^{(2)}$ are interesting since they correspond to the ODEs including a pseudo-differential operator, while generalizations to $B_{r}^{(1)}$, $A_{2r-1}^{(2)}$, $A_{2r}^{(2)}$, and $C_r^{(1)}$ is straightforward. In this section, we will diagonalize $D_{r}^{(1)}$-, $B_{r}^{(1)}$-, $D_{r+1}^{(2)}$-, and $A_{2r-1}^{(2)}$-type linear problems for lower-rank cases explicitly. The diagonalization can be applied directly except for $D_{r}^{(1)}$ where there are two sets of extra conserved densities. 

Let us begin with the modified Lax operator \eqref{eq:holo_mlax} for an affine Lie algebra $\hat{\mathfrak g}$:
\begin{equation}\label{eq:other_ml}
\epsilon\mathcal{L}_{m}=\epsilon\partial_{z}+\epsilon\sum_{i=1}^{r}\partial_{z}\phi_{i}(z)H_{i}+\sum_{i=1}^{r}E_{\alpha_{i}}+p(z)E_{\alpha_{0}}.
\end{equation}
where the representations for $H_{i}$ and $E_{\alpha_{i}}$ are defined in Appendix \ref{Appe.A}. 
After a series of gauge transformations, one obtains the diagonalized connection $A_{\rm diag}(z)$.
We denote $f(z,\epsilon)$ as the bottom component of diagonal elements of $A_{\rm diag}(z)$, which is expanded as
\begin{equation}\label{eq:diag_ele}
f(z,\epsilon)=\sum_{i=0}^{\infty}\epsilon^{i}f_{i}(z).
\end{equation}

In the last section, we have seen that the diagonal element $f(z,\epsilon)$ for $A_{r}^{(1)}$ is related to the WKB solution $P(z,\epsilon)$ to the adjoint ODE \eqref{eq:ad_Ar_ode}.
One of the goals of this section is to check the equality $f(z,\epsilon)=P(z,\epsilon)$ for other affine Lie algebras. For $B_{r}^{(1)}$ and $A_{2r-1}^{(2)}$ types, it is possible to find WKB solutions directly, and the equality can be checked. For $D_{r}^{(1)}$ and $D_{r+1}^{(2)}$ types, the diagonal elements may lead to the WKB solutions to the ODEs including a pseudo-differential operator, which will also be confirmed after taking specific limits. The adjoint ODEs of $D_{r}^{(1)},\,B_{r}^{(1)}$, $D_{r+1}^{(2)}$, and $A_{2r-1}^{(2)}$-type differential equations \eqref{eq:other_ode} are defined by $\partial_{z}\rightarrow-\partial_{z}$ and reversing the order of the action of differentials.

The form of $\sum_{i=1}^{r}\partial_{z}\phi_{i}(z,\Bar{z})H_{i}$ in Eq. \eqref{eq:other_ml} in our representation is a little bit tedious to write down. We make the following simplification:
\begin{align}
\begin{split}
&\quad\quad\quad\quad\quad\quad\quad\quad\quad\bm{\phi}_{i}= 
     \begin{cases}
       \phi_{1}, &\quad i=1,\\
        \phi_{i}-\phi_{i-1}, &\quad 2\leq i\leq r-2,\\
     \end{cases}\\
&\bm{\phi}_{r-1}= 
     \begin{cases}
       \phi_{r-1}+\phi_{r}-\phi_{r-2}, &\quad D_{r}^{(1)} ,\\
       \phi_{r-1}-\phi_{r-2}, &\quad \text{others},
    \end{cases}\quad
 \bm{\phi}_{r}= 
     \begin{cases}
       2\phi_{r}-\phi_{r-1}, &\quad B_{r}^{(1)},\; D_{r+1}^{(2)} ,\\
       \phi_{r}-\phi_{r-1}, &\quad \text{others}.
     \end{cases}
\end{split}
\end{align}
Define $D(\bm{\phi}_{i})\equiv\partial_{z}+\partial_{z}\bm{\phi}_{i}$ so that $\partial_z+\sum_{i=1}^{r}\partial_{z}\phi_{i}(z,\Bar{z})H_{i}$ in Eq. \eqref{eq:other_ml} and differential equations \eqref{eq:other_ode} can be rewritten in terms of $D(\bm{\phi}_{i})$. 

\subsection{Diagonalization of $A_{2r-1}^{(2)}$}
First, we begin with the WKB solution for the $A_{2r-1}^{(2)}$ linear problem in the $2r$-dimensional representation given in Appendix \ref{Appe.A}. $\Lambda$ in $2r$-dimensional representation is given by
\begin{equation}
\small
		\Lambda=\begin{pmatrix}
        0 & 1 &   &   &   \\
          & 0 & 1 &   &   \\
          &   & \ddots &  &   \\
        1 &   &   &   0     & 1 \\
          & 1  &   &    &  0\\ 
    \end{pmatrix},
\end{equation}
which has the eigenvalues: $\{0,1,e^{\frac{2\pi i }{h}},e^{\frac{4\pi i }{h}},\dots,e^{\frac{2\pi i (2r-2)}{h}}  \}$ with $h=2r-1$. The higher-order ordinary differential equation in Eq. \eqref{eq:other_ode} becomes
\begin{equation}\label{eq:ode_ar_2}
    \epsilon^{h}D(-\bm{\phi}_{1})\cdots D(-\bm{\phi}_{r})D(\bm{\phi}_{r})\cdots D(\bm{\phi}_{1})\psi(z,\epsilon)+2\sqrt{p(z)}\partial_{z}\sqrt{p(z)}\psi(z,\epsilon)=0.
\end{equation}

We apply the diagonalization of $A(z)$ from the bottom row with the gauge transformation $T_{2r}$. This implies the diagonalization conditions for $g_{2r,i}$'s: 
\begin{align}
{\bf Gau}_{T_{2r}}[A(z)]_{2r,1}=\cdots={\bf Gau}_{T_{2r}}[A(z)]_{2r,2r-1}=0.
\end{align}
${\bf Gau}_{T_{2r}}[A(z)]_{2r,i}=0$ $(2\leq i\leq 2r-1)$ eliminate $g_{2r,i}$ in terms of $g_{2r,2r-1}$.  Then ${\bf Gau}_{T_{2r}}[A(z)]_{2r,1}$ $=0$ determines the Riccati equation for $g_{2r,2r-1}$. 
Expanding $g_{2r,2r-1}=\sum_{i=0}^{\infty}(g_{2r,2r-1})_i\epsilon^i$, $(g_{2r,2r-1})_0$ satisfy
\begin{align}
    [(g_{2r,2r-1})_0(z)]^{2r}-2(g_{2r,2r-1})_{0}(z)p(z)=0.
\end{align}
Choosing a solution $(g_{2r,2r-1})_0=(2p)^{1/h}$, we can solve $g_{2r,2r-1}$ perturbatively in $\epsilon$.
The bottom diagonal component is given by
$f(z,\epsilon)=-g_{2r,2r-1}-\epsilon \bm{\phi}'_1$. 
The further diagonalization steps are similar and the final diagonal connection $A_{\text{diag}}(z)$ becomes
\begin{align}\label{eq:A2r-1_diag}
\textbf{Diag}\{h(z,\epsilon)+d(*), e^{-{2\pi i\over h}}f(z,e^{{2\pi i\over h}}\epsilon)+d(*),\dots, e^{-{2\pi i(h-1)\over h}}f(z,e^{{2\pi i(h-1)\over h}}\epsilon)+d(*),f(z,\epsilon)\},
\end{align}
where we fix $h(z,\epsilon)$ to be the top element determined by the traceless condition
\begin{equation}
    \sum_{n=0}^{2r-2}e^{-\frac{2\pi i n}{h}}f(z,e^{\frac{2\pi i n}{h}}\epsilon )+h(z,\epsilon)=d(*),
\end{equation}
and $f(z,\epsilon)$ obeys the Riccati equation for the adjoint of the ODE \eqref{eq:ode_ar_2}:
\begin{equation}\label{eq:ode_ar_2ad}
    \epsilon^{h}D(-\bm{\phi}_{1})\cdots D(-\bm{\phi}_{r})D(\bm{\phi}_{r})\cdots D(\bm{\phi}_{1})\psi(z,\epsilon)-2\sqrt{p(z)}\partial_{z}\sqrt{p(z)}\psi(z,\epsilon)=0
\end{equation}
with $\psi=\exp({1\over \epsilon}\int f(z,\epsilon)dz)$. The order of diagonal elements in Eq. \eqref{eq:A2r-1_diag} follows the convention in the $A_{r}^{(1)}$ type.

We now present the simplest example, the $A_{3}^{(2)}$ case.
The $f(z,\epsilon)$ for the first five orders are listed below
\begin{align}
    \begin{split}
        &f_{0}(z)=(2)^{\frac{1}{3}}p(z)^{\frac{1}{3}},\\
        &f_{1}(z)=-\frac{3}{2}\partial_{z}\ln f_{0},\\
        &f_{2}(z)=\frac{5 f_0''}{3 f_0{}^2}-\frac{5 f_{0}^{'2}}{2 f_0{}^2}+\frac{f_0''u_{2}}{3f_{0}},\\
        &f_{3}(z)=-\partial_{z}\big(\frac{u_2}{3 f_0^2}-\frac{5 \left(f_0'\right){}^2}{2 f_0^4}+\frac{5 f_0''}{3 f_0^3}\big),\\
        &f_{4}(z)=\frac{109 u_2 \left(f_0'\right){}^2}{36 f_0^5}-\frac{29 f_0' u_2'}{18 f_0^4}-\frac{17 u_2 f_0''}{18 f_0^4}+\frac{955 \left(f_0'\right){}^2 f_0''}{12 f_0^6}-\frac{955 \left(f_0'\right){}^4}{16 f_0^7}\\
        &\;\quad\quad-\frac{145 f_0{}^{(3)} f_0'}{9 f_0^5}-\frac{125 \left(f_0''\right){}^2}{12 f_0^5}+\frac{29 f_0{}^{(4)}}{18 f_0^4}-\frac{2u_{2}''+3u_{4}}{9f_0{}^3},\\
    \end{split}
\end{align}
with
\begin{align}
    \begin{split}
        &u_{2}(z)=\bm{\phi} _1'{}^2+\bm{\phi} _2'{}^2-3 \bm{\phi} _1''-\bm{\phi} _2'',\\
        &u_{4}(z)=\bm{\phi} _1'{}^2 \bm{\phi} _2''+2 \bm{\phi} _2' \bm{\phi} _1' \bm{\phi} _2''+\bm{\phi} _2'{}^2 \bm{\phi} _1''-\bm{\phi} _2'{}^2 \bm{\phi} _1'{}^2+\bm{\phi} _1{}^{(3)} \bm{\phi} _1'-\bm{\phi} _2{}^{(3)} \bm{\phi} _1'\\
        &\quad\quad\quad\; -\bm{\phi} _1'' \bm{\phi} _2''-\bm{\phi} _1{}^{(4)},
    \end{split}
\end{align}
and $h(z,\epsilon)=\sum_{i=0}^{\infty}\epsilon^{i}h_{i}(z)$ can be given by
\[   
h_{i}(z)= 
     \begin{cases}
       -3f_{i}(z), &\quad\text{if}\quad i=1+(2r-1)k,\; k\in \mathbf{Z},\\
       \quad 0, &\quad\text{otherwise.}\\
     \end{cases}
\]

\subsection{Diagonalization of $B_{r}^{(1)}$}
Next, we study the $B^{(1)}_r$ type linear problem in the $(2r+1)$-dimensional representation. From the Appendix \ref{Appe.A}, $\Lambda$ is given by
\begin{equation}
\small
		\Lambda=\begin{pmatrix}
			0 & 1 &  &  & & & & & &\\
			& 0 & 1 &  & & & & & &\\
			&  & \ddots &  & & & & & &\\
			& & & 0 & 1 & & & & &\\
			&  &  &  & 0 & \sqrt{2} &  & & &\\
			&  &  &  &  & 0 & \sqrt{2}& & &\\
			& & &  & &  &0 & 1 & &\\
			&  & & & &  &  &\ddots &  &\\
			 1&  & & & &  &  &  &0 & 1\\
			& 1& & & &  & & & & 0\\
		\end{pmatrix},
\end{equation}
whose eigenvalues are $2^{\frac{1}{r}}\{0,1, e^{\frac{2\pi i }{h}},e^{\frac{4\pi i }{h}},\dots,e^{\frac{2\pi i (2r-1)}{h}} \}$ with $h=2r$. The higher-order ordinary differential equation \eqref{eq:other_ode} obtained from the linear problem is
\begin{equation}\label{eq: Br ode}
\epsilon^{h}D(-\bm{\phi}_{1})\cdots D(-\bm{\phi}_{r})\partial_{z}D(\bm{\phi}_{r})\cdots D(\bm{\phi}_{1})\psi(z,\epsilon)-4\sqrt{p(z)}\partial_{z}\sqrt{p(z)}\psi(z,\epsilon)=0.
\end{equation}
The diagonalization approach can be applied without modification. ${\bf Gau}_{T_{2r+1}}[A(z)]_{2r+1,i}=0$ $(2\leq i\leq 2r)$ eliminate $g_{2r+1,i}$ in terms of $g_{2r+1,2r}$.  Then ${\bf Gau}_{T_{2r+2}}[A(z)]_{2r+1,1}=0$ determines the Riccati equation for $g_{2r+1,2r}$.
Expanding $g_{2r+1,2r}=\sum_{i=0}^{\infty}(g_{2r+1,2r})_i\epsilon^i$, $(g_{2r+1,2r})_0$ is shown to satisfy
\begin{align}
    [(g_{2r+1,2r})_0(z)]^{2r+1}-4(g_{2r+1,2r})_{0}(z)p(z)=0.
\end{align}
Here we will choose a solution $(g_{2r+1,2r})_0(z)=(4p(z))^{\frac{1}{h}}$. The bottom diagonal component is given by
$f(z,\epsilon)=-g_{2r+1,2r}-\epsilon \bm{\phi}'_1$ which is the solution to the Riccati equation of the ODE \eqref{eq: Br ode} with $\psi=\exp({1\over \epsilon}\int f(z,\epsilon)dz)$. 
Note that the ODE \eqref{eq: Br ode} is self-adjoint.
The diagonalized connection $A_{\text{diag}}$  can be summarized as
\begin{align}\label{eq:B2_diag}
\textbf{Diag}\{d(*), e^{-{2\pi i\over h}}f(z,e^{{2\pi i\over h}}\epsilon)+d(*),\dots, e^{-{2\pi i(2r-1)\over h}}f(z,e^{{2\pi i(2r-1)\over h}}\epsilon)+d(*),f(z,\epsilon)\},
\end{align}
where we choose the top element such that it corresponds to $0$ eigenvalue of $\Lambda$. The form \eqref{eq:B2_diag} has been checked for $r\leq4$. The traceless condition implies that $f_{1+2rk}(z)$ $(k\in\mathbf{Z})$ are total derivatives. 

Here are the first five nonzero terms for $B_{2}^{(1)}$ case.
\begin{align}
    \begin{split}
        &f_{0}(z)=-\sqrt{2}p(z)^{\frac{1}{4}},\\
        &f_{1}(z)=-2\partial_{z}\ln f_{0},\\
        &f_{2}(z)=-\frac{u_2}{4 f_0}-\frac{15 \left(f_0'\right){}^2}{4 f_0^3}+\frac{5 f_0''}{2 f_0^2},\\
        &f_{3}(z)=\partial_{z}\big(\frac{u_2}{4 f_0^2}+\frac{15 \left(f_0'\right){}^2}{4 f_0^4}-\frac{5 f_0''}{2f_0^3}\big),\\
        &f_{4}(z)=-\frac{33 u_2 \left(f_0'\right){}^2}{16 f_0^5}+\frac{9 f_0' u_2'}{8 f_0^4}+\frac{5 u_2 f_0''}{8 f_0^4}+\frac{885 \left(f_0'\right){}^2 f_0''}{8 f_0^6}\\
        &\;\quad\quad-\frac{2655 \left(f_0'\right){}^4}{32 f_0^7}-\frac{45 f_0{}^{(3)} f_0'}{2 f_0^5}-\frac{115 \left(f_0''\right){}^2}{8 f_0^5}+\frac{9 f_0{}^{(4)}}{4 f_0^4}+\frac{u_{2}^{2}+8u_{4}}{32f_0{}^3},\\
    \end{split}
\end{align}
with
\begin{align}
    \begin{split}
        &u_{2}(z)=-\bm{\phi} _1'{}^2-\bm{\phi} _2'{}^2+4\bm{\phi} _1''+2\bm{\phi} _2'',\\
        &u_{4}(z)=2 \bm{\phi} _1'{}^2 \bm{\phi} _2''+2 \bm{\phi} _2' \bm{\phi} _1' \bm{\phi} _2''+2 \bm{\phi} _2'{}^2 \bm{\phi} _1''-\bm{\phi} _2'{}^2 \bm{\phi} _1'{}^2+3 \bm{\phi} _1{}^{(3)} \bm{\phi} _1'-2 \bm{\phi} _2{}^{(3)} \bm{\phi} _1'\\
        &\quad\quad\quad\; +\bm{\phi} _2{}^{(3)} \bm{\phi} _2'+\bm{\phi} _2''{}^2-4 \bm{\phi} _1'' \bm{\phi} _2''-4 \bm{\phi} _1{}^{(4)}-\bm{\phi} _2{}^{(4)}.\\
    \end{split}
\end{align}
The results agree with the WKB solutions obtained by direct calculations.

\subsection{Diagonalization of $D_{r+1}^{(2)}$}
So far, we discussed the linear problem associated with the higher order ODE, including the differential operator $\partial_{z}$. For the $D$-type linear problem, we need to introduce the pseudo-differential operator $\partial_{z}^{-1}$ to reduce the linear problem to the single ODE, where we cannot apply the standard WKB approach. In the following two subsections, we will apply the diagonalization method to this problem. Let us first consider the $D_{r+1}^{(2)}$-type linear problem in the $(2r+2)$-dimensional representation. The $\Lambda$ is
\begin{equation}
\small
		\Lambda=\begin{pmatrix}
			0 & 1 &  &  & & & & & &\\
			& 0 & 1 &  & & & & & &\\
			&  & \ddots &  & & & & & &\\
			& & & 0 & \sqrt{2} & & & & &\\
			&  &  &  & 0 & & \sqrt{2} & & &\\
			\sqrt{2}&  &  &  &  & 0 & & & &\\
			& & &  & &  &0 & 1 & &\\
			&  & & & &  &  &\ddots &  &\\
			 &  & & & &  &  &  &0 & 1\\
			& & & & & \sqrt{2} & & & & 0\\
		\end{pmatrix}.
\end{equation}
The differential equation from the linear problem is also self-adjoint:
\begin{equation}\label{eq: Cr ode}
\epsilon^{h}D(-\bm{\phi}_{1})\cdots D(-\bm{\phi}_{r})\partial_{z}D(\bm{\phi}_{r})\cdots D(\bm{\phi}_{1})\psi(z,\epsilon)-4p(z)\partial^{-1}_{z}p(z)\psi(z,\epsilon)=0,
\end{equation}
with $h=2r+2$. We can apply the diagonalization method to the linear problem. The first step of the bottom-row diagonalization requires the conditions:
\begin{align}
{\bf Gau}_{T_{2r+2}}[A(z)]_{2r+2,1}=\cdots={\bf Gau}_{T_{2r+2}}[A(z)]_{2r+2,2r+1}=0.
\end{align}
${\bf Gau}_{T_{2r+2}}[A(z)]_{2r+2,i}=0$ $(2\leq i\leq 2r+1)$ eliminate $g_{2r+2,i}$ in terms of $g_{2r+2,2r+1}$. Then ${\bf Gau}_{T_{2r+2}}[A(z)]_{2r+2,1}=0$ determines the Riccati equation for $g_{2r+2,2r+1}$.
Expanding $g_{2r+2,2r+1}=\sum_{i=0}^{\infty}(g_{2r+2,2r+1})_i\epsilon^i$, $(g_{2r+2,2r+1})_0$ satisfy
\begin{align}
    [(g_{2r+2,2r+1})_0(z)]^{2r+2}-4p(z)^{2}=0.
\end{align}
Here we will choose a solution $(g_{2r+2,2r+1})_0(z)=(2p(z))^{\frac{1}{r+1}}$.
The bottom diagonal component is given by
$f(z,\epsilon)=-g_{2r+2,2r+1}-\epsilon \bm{\phi}'_1$. 
The further diagonalization steps are done in a similar way, and the final diagonal connection $A_{\text{diag}}(z)$ is given by
\begin{align}\label{eq: Cr dialx}
\textbf{Diag}\{e^{-{2\pi i\over h}}f(z,e^{{2\pi i\over h}}\epsilon)+d(*),\dots, e^{-{2\pi i(2r+1)\over h}}f(z,e^{{2\pi i(2r+1)\over h}}\epsilon)+d(*),f(z,\epsilon)\}.
\end{align}
The traceless condition implies the $f_{1+(2r+2)k}(z)$ $(k\in\mathbf{Z})$ are total derivatives. 

We discuss the simplest example $D_3^{(2)}$. The first five non-total derivative terms for $D_{3}^{(2)}$ are
\begin{align}
    \begin{split}
        &f_{0}(z)=-\big(2p(z)\big)^{\frac{1}{3}},\\
        &f_{1}(z)=-2\partial_{z}\ln f_{0},\\
        &f_{2}(z)=-\frac{u_2}{6 f_0}-\frac{5 \left(f_0'\right){}^2}{2 f_0^3}+\frac{5 f_0''}{3 f_0^2},\\
        &f_{3}(z)=0,\\
        &f_{4}(z)=\frac{61 u_2 \left(f_0'\right){}^2}{36 f_0^5}-\frac{7 f_0' u_2'}{9 f_0^4}-\frac{11 u_2 f_0''}{18 f_0^4}-\frac{475 \left(f_0'\right){}^2 f_0''}{6 f_0^6}\\
        &\;\quad\quad+\frac{475 \left(f_0'\right){}^4}{8 f_0^7}+\frac{140 f_0{}^{(3)} f_0'}{9 f_0^5}+\frac{65 \left(f_0''\right){}^2}{6 f_0^5}-\frac{14 f_0{}^{(4)}}{9 f_0^4}+\frac{3u_{2}^{2}+12u_{4}+22u_{2}''}{72 S_{0}^{4}},
    \end{split}
\end{align}
with
\begin{align}
    \begin{split}
        &u_{2}(z)=-\bm{\phi} _1'{}^2-\bm{\phi} _2'{}^2+4\bm{\phi} _1''+2\bm{\phi} _2'', \\
        &u_{4}(z)=2 \bm{\phi} _1'{}^2 \bm{\phi} _2''+2 \bm{\phi} _2' \bm{\phi} _1' \bm{\phi} _2''+2 \bm{\phi} _2'{}^2 \bm{\phi} _1''-\bm{\phi} _2'{}^2 \bm{\phi} _1'{}^2+3 \bm{\phi} _1{}^{(3)} \bm{\phi} _1'-2 \bm{\phi} _2{}^{(3)} \bm{\phi} _1'\\
        &\quad\quad\quad\; +\bm{\phi} _2{}^{(3)} \bm{\phi} _2'+\bm{\phi} _2''{}^2-4 \bm{\phi} _1'' \bm{\phi} _2''-4 \bm{\phi} _1{}^{(4)}-\bm{\phi} _2{}^{(4)}. \\
    \end{split}
\end{align}
The perturbative results for $f(z,\epsilon)$ here agree with those for $P_{1}(z,\epsilon)$ in the case of $\phi_{i}(z)=0$. See Appendix \ref{Appe.B}.  

\subsection{Diagonalization of $D_{r}^{(1)}$}
Finally, we study $D_{r}^{(1)}$ case. 
Here we consider the linear problem in the $2r$-dimensional vector representation of $D_r$. 
The matrix $\Lambda$ in Eq. \eqref{eq:toda_lax} is given by
\begin{equation}
\small
		\Lambda=\begin{pmatrix}
			0 & 1 &  &  & & & & & &\\
			& 0 & 1 &  & & & & & &\\
			& & \ddots & & & & & & &\\
			&  &  & 0 & 1 & 1 & & & &\\
			&  &  &  & 0 & &1 & & &\\
			& & & &  &0 &1 & & &\\
			& &  & & & & 0 & 1 & &\\
			& &  & & & & &\ddots & &\\
			1&  & & & & &  &  & 0 & 1\\
			&1 & & &  &  &  &  &  & 0\\
		\end{pmatrix},
\end{equation}
with the eigenvalues: $2^{\frac{1}{r-1}}\{0,0,1, e^{\frac{2\pi i }{h}},e^{\frac{4\pi i }{h}},\dots,e^{\frac{2\pi i (h-1)}{h}} \}$. Due to the existence of double zero eigenvalues in $\Lambda$, we introduce a matrix $F$ commuting with $\Lambda$, which is given by
\begin{equation}\label{}
    F=ie_{1,r}-ie_{1,r+1}-ie_{r,1}+ie_{r+1,1}+ie_{r,2r}-ie_{r+1,2r}-ie_{2r,r}+ie_{2r,r+1}.
\end{equation}
The matrix $F$ with eigenvalues $\{-2,2,0,0,0,\dots,0\}$ is nothing but the one that appears in Eq. \eqref{eq:h2}. It leads to a new set of conserved densities. The corresponding ODE associated with the linear model is given by \cite{Ito:2013aea}. It takes the form of
\begin{equation}\label{eq:Dr_ode}
\epsilon^{h}D(-\bm{\phi}_{1})\cdots D(-\bm{\phi}_{r})\partial_{z}^{-1}D(\bm{\phi}_{r})\cdots D(\bm{\phi}_{1})\psi(z,\epsilon)-4\sqrt{p(z)}\partial_{z}\sqrt{p(z)}\psi(z,\epsilon)=0,
\end{equation}
with $h=2r-2$. This equation includes the pseudo-differential operator.

We apply the same diagonalization steps for $A(z)$ from the bottom row with the following conditions for $g_{2r,i}$: 
\begin{align}
{\bf Gau}_{T_{2r}}[A(z)]_{2r,1}=\cdots={\bf Gau}_{T_{2r}}[A(z)]_{2r,2r-1}=0.
\end{align}
These equations, however, cannot be expressed locally in terms of $g_{2r,2r-1}$ in contrast to our previous examples. The most concise way to solve the constraints is to express $g_{2r,i}$ in terms of $g_{2r,2r-1}$ and $g_{2r,r+1}$, which give rise to the two Riccati equations:
\begin{align}\label{eq:riccati_ar_1}
{\bf Gau}_{T_{2r}}[A(z)]_{2r,1}&=0, \quad {\bf Gau}_{T_{2r}}[A(z)]_{2r,r}=0.
\end{align}
Substituting the expansions $g_{2r,2r-1}=\sum_{i=0}^{\infty}(g_{2r,2r-1})_i \epsilon^i$ and 
$g_{2r,r+1}=\sum_{i=0}^{\infty}(g_{2r,r+1})_i \epsilon^i$ to \eqref{eq:riccati_ar_1},
one can solve   Eqs. \eqref{eq:riccati_ar_1}  recursively in $\epsilon$.
Here $(g_{2r,2r-1})_0$ and $(g_{2r,r+1})_0$ satisfy 
\begin{align}
2 (g_{2r,r+1})_0 (g_{2r,2r-1})_0-[(g_{2r,2r-1})_0]^r&=0, \quad -4p(z) (g_{2r,2r-1})_0+(g_{2r,r+1})_0 [(g_{2r,2r-1})_0]^r=0.
\end{align}
Then $(g_{2r,2r-1})_0$ satisfies $(g_{2r,2r-1})_0=0$ or $4p(z)=[(g_{2r,2r-1})_0]^h$.
Solving Eqs. \eqref{eq:riccati_ar_1}, one obtains the bottom diagonal element $f(z,\epsilon)=-g_{2r,2r-1}-\epsilon\phi'_1$. The $i$-th row elements can be diagonalized similarly for $r+2\leq i\leq 2r$. For $i\leq r+1$, the procedure can be reduced to solve one single Riccati equation for $g_{i,i-1}$.

We have performed the diagonalization procedure and obtained  $A_{\text{diag}}(z)$ for $r\leq4$, where we have chosen $(g_{2r,2r-1})_0=(4 p(z))^{1/h}$.
The diagonalized connection can be summarized as
\begin{align}
    \begin{split}
        &\textbf{Diag}\{e^{\frac{-2\pi i}{h}}f(z,e^{\frac{2\pi i}{h}}\epsilon)+d(*),\dots, e^{\frac{-2\pi i(r-1)}{h}}f(z,e^{\frac{2\pi i(r-1)}{h}}\epsilon)+d(*),\, e^{i\pi (r-1)}K(z,-\epsilon)+d(*),\\
        &K(z,\epsilon),\, e^{-\frac{2\pi i r}{h}}f(z,e^{\frac{2\pi i r}{h}}\epsilon)+d(*),\dots, e^{-\frac{2\pi i (2r-1)}{h}}f(z,e^{\frac{2\pi i (2r-1)}{h}}\epsilon)+d(*),\, f(z,\epsilon) \}.\\
    \end{split}
\end{align}
The elements $K(z,\epsilon)$ (resp. $K(z,-\epsilon)$) corresponding to eigenvalues $\pm2$ of $F$ are obtained from the $(r+1)$ (resp. $r$)-th diagonalization conditions
\begin{align}
\begin{split}
&{\bf Gau}_{T_{r+1}\dots T_{2r}}[A(z)]_{r+1,i}=0, \; 1\leq i\leq 2r, \; i\neq r+1, \\
&{\bf Gau}_{T_{r}\dots T_{2r}}[A(z)]_{r,i}=0, \; 1\leq i\leq 2r, \; i\neq r.\\
\end{split}
\end{align}
${\bf Gau}_{T_{r+1}\dots T_{2r}}[A(z)]_{r+1,1}=0$ (resp. ${\bf Gau}_{T_{r}\dots T_{2r}}[A(z)]_{r,1}=0$) determines the Riccati equations for $g_{r+1,r}$ (resp. $g_{r.r-1}$). $K(z,\epsilon)$ (resp. $e^{-\frac{2\pi i r}{h}}K(z,-\epsilon)$) are namely the $r+1$ (resp. $r$)-th diagonal elements in terms of $g_{r+1,r}$($g_{r.r-1}$) up to total derivatives. The expansion of $K(z,\epsilon)=\sum_{i=0}^{\infty}\epsilon^{i}K_{i}(z)$ corresponds to the $J_{i}(z)$ terms in Eq. \eqref{eq:h2}. Note that the traceless condition of $A_{\rm diag}$ implies the $f_{1+(2r-2)k}(z)$ $(k\in\mathbf{Z})$ are total derivatives. 

We now give the simplest example $D_3^{(1)}$ in the six-dimensional representation.
This is equivalent to the $A_3^{(1)}$ in the same dimensional representation.
The first five nonzero terms of $D_{3}^{(1)}$ are given by
\begin{align}
    \begin{split}
        &f_{0}(z)=-\sqrt{2}p(z)^{\frac{1}{4}},\\
        &f_{1}(z)=-2\partial_{z}\ln f_{0},\\
        &f_{2}(z)=-\frac{u_2}{4 f_0}-\frac{15 \left(f_0'\right){}^2}{4 f_0^3}+\frac{5 f_0''}{2 f_0^2},\\
        &f_{3}(z)=\partial_{z}\big(\frac{u_2}{4 f_0^2}+\frac{15 \left(f_0'\right){}^2}{4 f_0^4}-\frac{5 f_0''}{2f_0^3}\big),\\
        &f_{4}(z)=-\frac{33 u_2 \left(f_0'\right){}^2}{16 f_0^5}+\frac{9 f_0' u_2'}{8 f_0^4}+\frac{5 u_2 f_0''}{8 f_0^4}+\frac{885 \left(f_0'\right){}^2 f_0''}{8 f_0^6}\\
        &\;\quad\quad-\frac{2655 \left(f_0'\right){}^4}{32 f_0^7}-\frac{45 f_0{}^{(3)} f_0'}{2 f_0^5}-\frac{115 \left(f_0''\right){}^2}{8 f_0^5}+\frac{9 f_0{}^{(4)}}{4 f_0^4}+\frac{u_{2}^{2}+8u_{4}}{32f_0{}^3},\\
        &K_{3}(z)=\frac{v_{3}}{f_{0}{}^{2}},
    \end{split}
\end{align}
with
\begin{align}
    \begin{split}
        &u_{2}(z)=-\bm{\phi} _1'{}^2-\bm{\phi} _2'{}^2-\bm{\phi} _3'{}^2+4\bm{\phi} _1''+2\bm{\phi} _2'',\\
        &u_{4}(z)=2 \bm{\phi} _1'{}^2 \bm{\phi} _2''+2 \bm{\phi} _2' \bm{\phi} _1' \bm{\phi} _2''+2 \bm{\phi} _3' \bm{\phi} _1' \bm{\phi} _3''+2 \bm{\phi} _2'{}^2 \bm{\phi} _1''+2 \bm{\phi} _3'{}^2 \bm{\phi} _1''+2 \bm{\phi} _2' \bm{\phi} _3' \bm{\phi} _3''\\
        &\quad\quad\quad\; -\bm{\phi} _3'{}^2 \bm{\phi} _1'{}^2 +3 \bm{\phi} _1{}^{(3)} \bm{\phi} _1'-2 \bm{\phi} _2{}^{(3)} \bm{\phi} _1'-\bm{\phi} _1'{}^2 \bm{\phi} _2'{}^2-\bm{\phi} _2'{}^2 \bm{\phi} _3'{}^2+\bm{\phi} _2{}^{(3)} \bm{\phi} _2'\\
        &\quad\quad\quad\; +\bm{\phi} _3{}^{(3)} \bm{\phi} _3'+\bm{\phi} _2''{}^2 -4 \bm{\phi} _1'' \bm{\phi} _2''-4 \bm{\phi} _1{}^{(4)}-\bm{\phi} _2{}^{(4)},\\
        &v_{3}(z)=\bm{\phi} _3' \bm{\phi} _2''+\bm{\phi} _1' \bm{\phi} _3''+\bm{\phi} _2' \bm{\phi} _3''-\bm{\phi} _1' \bm{\phi} _2' \bm{\phi} _3'-\bm{\phi} _3{}^{(3)},
    \end{split}
\end{align}
where $K_{3}(z)$ vanishes when $\bm{\phi}_{3}=0$. It can be understood from the fact that the pseudo-differential operator is canceled in Eq. \eqref{eq:Dr_ode}, and the ODE is reduced to $B_{r-1}^{(1)}$ type for $\bm{\phi}_{3}=0$. The diagonalized connection obtained here confirms this relation. One may notice the diagonal elements share the same form between $B_{r-1}^{(1)}$ and $D_{r}^{(1)}$ types and the difference only appears in the construction of $u_{i}$.

\subsection{Remarks on $f(z,\epsilon)$}
The diagonal element $f(z,\epsilon)$ in $A_{\rm diag}(z)$ for $D_{r}^{(1)},\,B_{r}^{(1)}$, $D_{r+1}^{(2)}$ and $\,A_{2r-1}^{(2)}$ shares the following common properties.

First, during the diagonalization of the linear problem for an affine Lie algebra $\hat{\mathfrak g}$ except $A_r^{(1)}$, we observe that all of $f_{2i-1}(z)$ $(i\in\mathbb{N})$ are total derivatives. It is explained by Eq. \eqref{eq: DS dlax gr} where $H(z,\epsilon)$ is expanded in terms of $\Lambda^{-(2i-1)}$ ($i\in {\mathbb N}$) and the coefficients in $\Lambda^{-2i}$ are total derivatives. However, this is not true for $A_{r}^{(1)}$ type, where Eq. \eqref{eq: DS dlax} admits the even terms $\Lambda^{-2i}$, and $f_{2i-1}(z)$ are non-zero in general. 

Second, we can consider the $\phi_i=0$ case, where $D(\phi_i)$ becomes the differential operator $\partial$, and the ODE becomes simple.
In this case, it is possible to study higher-order corrections for an affine Lie algebra with higher ranks.
The diagonal elements $f_i(z)$ are uniquely determined up to total derivatives. 
For the practical computation of the WKB periods, where the total derivatives are not relevant,  it is convenient to express them in a simple form.
It is possible to make the following simplification:
\begin{equation}
    f_{i}(f_{0},f_{0}',f_{0}'',\cdots)=S_{i}(f_{0},f_{0}'',\cdots)+d(*),
\end{equation}
where $S_{i}$ is independent of $f_{0}'$. All the $f_{0}'$ terms are absorbed into total derivatives $d(*)$.
We found that $S_{i}(z)$ can be fixed by $S_{0}=f_{0}$. The details for $A_r^{(1)}$, $B_r^{(1)}$, $D^{(1)}_r$, $D^{(2)}_{r+1}$, and $A^{(2)}_{2r-1}$ are summarized in Appendix \ref{Appe.c}.

\section{Conserved densities and WKB solutions}\label{Sec_5}
In Sections \ref{sect:Sec3} and \ref{sect:Sec_4}, we have studied the WKB solutions of the linear problem for a classical affine Lie algebra $\hat{\mathfrak g}$ from diagonalizing the connection $A(z)$. The diagonalized connection $A_{\textbf{diag}}(z)$ has a structure similar to the conserved densities which appear in the construction of the generalized KdV hierarchies associated with $\hat{\mathfrak g}$. We will see the correspondence between the coefficients in the WKB expansion and the conserved densities for the $A_r^{(1)}$-type linear problem.  
So far, we have checked the correspondence above for several affine Lie algebras with low ranks. Based on the observations, we conjecture it holds for other affine Lie algebras, which is left for future works.

\subsection{Continuity equations from the KdV hierarchies}
First, let us review the generalized KdV hierarchies associated with the Lax operator ${\cal L}$ in Eq. \eqref{eq:toda_lax} for an affine Lie algebra $A_r^{(1)}$  \cite{Drinfeld:1984qv}. The adjoint ordinary differential equation obtained from the linear problem is given in Eq. \eqref{eq:ad_Ar_ode}, where the operator on the left-hand side is nothing but the scalar Lax operator for the modified KdV hierarchies.
\begin{equation}\label{eq:dslax}
    L_{\text{scalar}}=(\partial_{x}-\partial_{x}\phi_{1})\cdots(\partial_{x}+\partial_{x}\phi_{2}-\partial_{x}\phi_{1})(\partial_{x}+\partial_{x}\phi_{r}),
\end{equation}
where we have changed the coordinates ($z, \bar{z}$) into $(x,t)$. The coordinates $x$ and $t$ are real in this setup.  The operator \eqref{eq:dslax} can be written in the canonical form:
\begin{equation}\label{op:cslax}
    L_{\text{can}}=\partial^{r+1}_{x}-\sum_{i=0}^{r-1}u_{r+1-i}(x)\partial^{i}_{x}.
\end{equation}
Here $u_i(x)$ is expressed in terms of $\phi_i$'s obtained by the Miura transformation. In the $A_1^{(1)}$ case, it leads to
\begin{equation}
    u_{2}(x)=(\partial_{x}\phi_{1})^{2}-\partial^{2}_{x}\phi_{1},
\end{equation}
which is the classical energy-momentum tensor in the sinh-Gordon model \cite{Drinfeld:1984qv, Bazhanov:1994ft}. The equivalence between ${\cal L}$ in Eq. \eqref{eq:toda_lax} and $L_{\rm can}$ has been proved in \cite{Drinfeld:1984qv}. Both the Lie algebraic and the scalar Lax operators satisfy the Lax equation. Let us first focus on the scalar one. The linear problem can be written as the equation $L\psi(x)=\lambda\psi(x)$ with $\lambda$ the eigenvalue of $L$. Introducing  time parameters $t_i$ ($i=1,2,\dots)$ with $t_1=t$,  the integrable hierarchies are defined by the Lax equation
\begin{equation}
    \partial_{t_{i}}L=[A_{i},L],
\end{equation}
with $A_{i}=(L^{\frac{i}{h}})_{+}$, where $(A)_+$ denotes the non-negative part in $\partial_{x}$ of the differential operator $A$.   Further acting $\partial_{t_{i}}$ on $L\psi(x)=\lambda\psi(x)$, one can obtain $(L-\lambda)(\partial_{t_{i}}\psi-A_{i}\psi)=0$ which implies, for some function $g(t_{i})$,
\begin{equation}
\partial_{t_{i}}\psi(x)-A_{i}\psi(x)=g(t_{i})\psi(x).
\end{equation}
Substitute the WKB expansion $\psi(x,\epsilon)=\exp(\frac{1}{\epsilon}\int dx\,P(x,\epsilon))$, it leads to
\begin{equation}
    \partial_{t_{i}}P(x,\epsilon)-\partial_{x}a_{i}(x)=0,
\end{equation}
with $a_{i}(x,\epsilon)=\epsilon A_{i}\psi(x,\epsilon)/\psi(x,\epsilon)$. This equation implies $P(x,\epsilon)$ is the conserved density. 

On the other hand, the integrable hierarchies defined by ${\cal L}$ are given by
\begin{equation}
    \partial_{t}\mathcal{L}=[\mathcal{A},\mathcal{L}],
\end{equation}
where $\mathcal{A}(x)=\sum_{i=0}^{\infty}\mathcal{A}_{i}(x)(\lambda\Lambda)^{-i}$. After the diagonalization, the Lax operator ${\cal L}_{\text{diag}}$ is the form of Eq. \eqref{eq:DS_dilax}. One can obtain $\partial_{t}\mathcal{L}_{\textbf{diag}}=[\mathcal{A}',\mathcal{L}_{\textbf{diag}}]$ with $\mathcal{A}'=T^{-1}\mathcal{A}T$. This is nothing but the continuity equation
\begin{equation}
    \partial_{t}f_{i}+\partial_{x}\mathcal{A}'_{i}=0.
\end{equation}
The functions $f_{i}(x)$ are also conserved densities. Since both $P(x,\epsilon)$ and the diagonal element $f(x,\epsilon)$ are conserved densities, it implies the equality $f(x,\epsilon)=P(x,\epsilon)$ up to total derivatives.

\subsection{Conserved densities in $A_{1}^{(1)}$ and $A_{2}^{(1)}$ affine Toda field theories}
We have seen the correspondence between the WKB expansions and the conserved densities for $A_r^{(1)}$ with $p(z)=1$ case. 
Here we show the relations between WKB solutions and classical conserved densities with the conformal transformation. It is convenient to begin with the canonical Lax operator \eqref{eq: mcan Lax}. Recall the appearance of $p(z)$: the conformal transformation Eq. \eqref{eq: cft}. It implies \cite{Lukyanov:2010rn}
\begin{equation}\label{eq: ucft}
    dw=\sqrt{p(z)}dz, \quad \hat{u}_{2}\big(w(z)\big)=\frac{1}{p(z)}\Big[ u_{2}(z)+\frac{4pp''-5p'^{2}}{16p^{2}}\Big],
\end{equation}
with $u_{2}(z)=\phi'(z)^2-\phi''(z)$. Then the modified canonical Lax operator \eqref{eq: mcan Lax} for $A_{1}^{(1)}$ in the representation can be written as
\begin{equation}
\mathcal{L}_{\text{can}}=\epsilon\partial_{w}+e_{1,2}+\big(\epsilon^{2}\hat{u}_{2}(w)+1\big)e_{2,1}.
\end{equation}
After the similar diagonalization procedure in Section \ref{sec_3.3}, one can see the bottom diagonal element satisfies
\begin{equation}\label{eq: A11 cRacatti}
    \hat{f}(w,\epsilon)^2+\epsilon \partial_{w}\hat{f}(w,\epsilon)-\epsilon ^2 \hat{u}_{2}(w)-1=0.
\end{equation}
The expansion $\hat{f}(w,\epsilon)=\sum_{i=0}^{\infty}\epsilon^{i}f_{i}(w)$ can be calculated perturbatively. The first five terms are 
\begin{align}
\begin{split}
        \hat{f}_{0}(w)&=1,\\
        \hat{f}_{1}(w)&=0,\\
        \hat{f}_{2}(w)&=\frac{\hat{u}_{2}(w)}{2},\\
        \hat{f}_{3}(w)&=-\frac{\partial_{w}\hat{u}_{2}(w)}{4},\\
        \hat{f}_{4}(w)&=\frac{\partial_{w}^{2}\hat{u}_{2}(w)-\hat{u}_{2}^{2}(w)}{8},\\
\end{split}
\end{align}
which are the conserved densities up to total derivatives \cite{Lukyanov:2010rn, Babelon:2003qtg, Bazhanov:1994ft}.\footnote{One may need $u_{0}\rightarrow -u_{0}$ according to different convention in $L=\partial^{2}\pm u$.} Substitute $\hat{u}(w)$ with Eq. \eqref{eq: ucft} and multiply $\sqrt{p(z)}$, one can obtain the solution to Eq. \eqref{eq: A11 mono Ricatti} again. $\sqrt{p(z)}$ is from coordinate transformation in integral of motions: $\oint dw\; \hat{f}_{i}\rightarrow \oint dz\; \sqrt{p(z)}f_{i}$. In conclusion, $\Pi_{i}$ defined below and conserved charges $\mathcal{Q}_{i}$ are related as follows:
\begin{equation}
    \Pi_{i}\equiv \oint dz\; f_{i}(z)=\oint dz\; \sqrt{p(z)}\hat{f}_{i}=\oint dw\; \hat{f}_{i}\equiv \mathcal{Q}_{i}.
\end{equation}
This implies that the WKB periods of the linear problem for an affine Toda field theory is a generating function of the classical conserved charges of the integrable hierarchies. 
It is known that for an affine Toda lattice, a finite-dimensional version of the present model, the integrals of motion are given by the period integral over the spectral curves. $\sum_{i=0}^{\infty}\epsilon^{i}\Pi_{i}$ defined on the spectral curve can be viewed as the quantum Seiberg-Witten periods in $4d$ $\mathcal{N}=2$ super Yang-Mills theories (SYMs). The equality here can be another evidence of the correspondence between SYMs and integrable models  \cite{Gorsky:1995zq,Martinec:1995by,Donagi:1995cf,DHoker:1999yni}. See also \cite{Fioravanti:2019vxi} for the relation to the Gelfand-Dickii algebra.

Similar results can also be obtained from the $A_{2}^{(1)}$ affine Toda field equation. The generalized Miura transformation:
\begin{equation}
    (\partial_{z}-\partial_{z}\phi_{1})(\partial_{z}-\partial_{z}\phi_{2}+\partial_{z}\phi_{1})(\partial_{z}+\partial_{z}\phi_{2})=\partial_{z}^{3}-\sum_{i=0}^{2}u_{3-i}\partial_{z}^{i}
\end{equation}
leads to the Miura transformation in Eq. \eqref{eq:A2miura}. After the transformation \cite{Mathieu:1988pm}: $u_{2}(z)\rightarrow 2u_{2}(z)$, $u_{3}(z)\rightarrow u_{3}(z)+u_{2}'(z)$, $u_{1}(z)$ becomes the classical energy-momentum tensor and $u_{3}(z)$ becomes the additional classical local spin-3 field in $\mathcal{W}_{3}$ conformal field theory \cite{Bazhanov:2001xm}. The conformal transformation leads to
\begin{equation}\label{eq: A21 cft}
    p(z)=(\partial_{z}w)^{3},\quad \hat{\phi}_{i}(w)=\phi_{i}(z)-\frac{1}{3}\log p(z).
\end{equation}
The modified canonical Lax operator now becomes
\begin{equation}
\mathcal{L}_{\text{can}}=\epsilon\partial_{w}+\epsilon^{2}u_{2}(z)e_{2,1}+(\epsilon^{3}u_{3}(z)+1)e_{3,1}+e_{1,2}+e_{2,3}.
\end{equation}
The first diagonalization condition gives the Riccati equation of $\hat{f}(w,\epsilon)$:
\begin{equation}\label{eq: A2 riccati}
    \hat{f}^3+3 \epsilon  \hat{f} \hat{f}'-\epsilon ^{2}\hat{u}_{2}\hat{f}+\epsilon ^{2}\hat{f}'' -\epsilon^3 \hat{u}_{3}-1=0,
\end{equation}
and the second diagonalization condition only leads to phase rotation. So the diagonal elements can be summarized by
\begin{equation}
    \hat{A}(w)=\textbf{Diag}\{e^{-\frac{2\pi}{3}}\hat{f}(w,e^{\frac{2\pi}{3}}\epsilon)+d(*), e^{-\frac{4\pi}{3}}\hat{f}(w,e^{\frac{4\pi}{3}}\epsilon)+d(*), \hat{f}(w,\epsilon)\}.
\end{equation}
The first five perturbative solutions to Eq. \eqref{eq: A2 riccati} are nothing but the conserved densities up to total derivatives
\begin{align*}
        \hat{f}_{0}(w)&=1,\\
        \hat{f}_{1}(w)&=0,\\
        \hat{f}_{2}(w)&=\frac{\hat{u}_{2}(w)}{3},\\
        \hat{f}_{3}(w)&=\frac{\hat{u}_{3}(w)-\partial_{w}\hat{u}_2(w)}{3},\\
        \hat{f}_{4}(w)&=\frac{1}{9}\left(3 \partial_{w}u_{3}(w)-\partial_{w}^{2}u_2(w)\right),\\
        \hat{f}_{5}(w)&=\frac{1}{9} \left(\hat{u}_3(w) \hat{u}_2(w)-2 \partial_{w}^{2}\hat{u}_3(w)+\partial_{w}^{3}\hat{u}_2{}(w)\right),
\end{align*}
where $\hat{f}_{4}(w)$ is a total derivative. Substitute the classical conformal transformation Eq. \eqref{eq: A21 cft} and multiply $\big(p(z)\big)^{\frac{1}{3}}$, one can obtain the same WKB solution to Eq. \eqref{eq:A21_ode}.

\section{Conclusions and Discussion}
In this paper, we have studied the WKB expansion of the linear problem for the modified affine Toda field equations after diagonalizing the connection with gauge transformation. 
We have obtained the diagonalization conditions for the gauge transformation parameters, which take the 
form of the Riccati equation for the WKB solution to the adjoint ODE of the linear problem.  
Solving the Riccati equation, we obtained the bottom component of the diagonalized connection.
After eliminating the total derivatives appropriately, other elements of the diagonalized connection are related by the ${\mathbb Z}_h$ phase rotation and the traceless condition \cite{Ito:2021boh}. 
For the D-type affine Lie algebras, there exist extra elements related to the conserved densities.
We have checked the results for $A_{r}^{(1)}$, $B_{r}^{(1)}$, $D_{r+1}^{(2)}$, $A_{2r+1}^{(2)}$ types and $D_{r}^{(1)}$ types in the specific limit with lower ranks. 
The present diagonalization approach provides a new method to find the WKB solution of the ordinary differential equations including a pseudo-differential operator.
We can apply this method to find the WKB solution to the linear problem for the affine Lie algebras of exceptional types, which provides an interesting future problem.
In the $A_{1}^{(1)}$ and $A_{2}^{(1)}$ models, we confirmed that the diagonal elements are classical conserved densities after conformal transformation, which should be valid for other affine Lie algebras. 

After taking the light-cone and conformal limits, we obtain the higher-order ODEs associated with the
modified affine Toda field equations.
The exact WKB periods for the ODEs are related to the Voros symbols, which are also regarded as the Y-functions and the cluster variables in the quantum integrable models \cite{iwaki2014exact}. 
It is interesting to study the WKB periods based on the Y-system and the related TBA equations. 
We also find the correspondence between the classical conserved densities and the WKB periods, which implies the equivalence between the classical and quantum conserved charges as in \cite{Bazhanov:1994ft, Bazhanov:2001xm, Feigin:2007mr,Masoero:2018rel}.
It is interesting to study the correspondence for general affine Lie algebras. 
The present class of ODEs also appears in the context of quantum Seiberg-Witten curves \cite{Ito:2017ypt}. In particular, it applies to the quantum periods of ${\cal N}=2$ super Yang-Mills theory based on a simple Lie group. The resurgence structure of the quantum SW periods would be helpful to understand the strong coupling physics of the theory.
 
 Furthermore, it is also interesting to generalize the diagonalization approach to supersymmetric integrable models. The $\mathcal{N}=1$ super sine-Gordon equations lead to the Schr\"{o}dinger equation with quadratic potential \cite{Ito:2022cev}. However, the integral of motions contributed by fermion is absent in the ODE. If one can diagonalize the $\mathcal{N}=1$ super linear problem, it may provide a new way to investigate 2D super integrable models. Finally, it has been shown that the classical $T\Bar{T}$ deformation can be viewed as a coordinate transformation \cite{Aramini:2022wbn}. One can combine the diagonalization and $T\Bar{T}$-deformation for integrable models. 

\section*{Acknowledgments}
We would like to thank C. Wu for useful explanations for $D$-type linear problems. 
We are also grateful to Georg\H{o} Nemes for the useful discussion.
The work of K.I. is supported in part by Grant-in-Aid for Scientific Research 21K03570 from Japan Society for the Promotion of Science (JSPS).
\appendix

\section{The representations for affine Lie algebras}\label{Appe.A}
In this appendix,  we show the representations for the generators of affine Lie algebras $\hat{\mathfrak{g}}$ used in this paper. The parameter $r$ and $h$ are the rank and the Coxeter number of the simple Lie algebra $\mathfrak{g}$. $e_{i,j}$ below denotes the matrix with components $(e_{i,j})_{a,b}=\delta_{i,a}\delta_{j,b}$. $\alpha_{i}$ $(1\leq i\leq r)$ denote the simple roots and $\alpha_{0}$ denotes the affine root. We also set $E_{-\alpha_{i}}=E_{\alpha_{i}}^{\text{T}}$. 
\begin{description}
\item[$A_r^{(1)}$:]
The Coxeter number $h=r+1$.
The $(r+1)$-dimensional representation is ($r \ge 1$)
\begin{gather}
  E_{\alpha_0} = e_{r+1,1}, \quad E_{\alpha_i} = e_{i,i+1} .
\end{gather}

\item[$B_r^{(1)}$:]
The Coxeter number $h=2r$.
The $(2r+1)$-dimensional representation is ($r \ge 2$)
\begin{gather}
  E_{\alpha_0} = e_{2r,1} + e_{2r+1,2} , \quad  E_{\alpha_i} = e_{i,i+1} + e_{2r+1-i,2r+2-i} , \notag\\
  E_{\alpha_r} = \sqrt{2}(e_{r,r+1} + e_{r+1,r+2}) .
\end{gather}

\item[$D_r^{(1)}$:]

The Coxeter number $h=2r-2$.
The $2r$-dimensional representation is ($r \ge 3$)
\begin{gather}
  E_{\alpha_0} = e_{2r-1,1}+e_{2r,2}, \quad E_{\alpha_i} = e_{i,i+1}+e_{2r-i,2r+1-i}, \quad E_{\alpha_r} = e_{r-1,r+1}+e_{r,r+2}.
\end{gather}

\item[$A_{2r-1}^{(2)}$:]

The Coxeter number $h=2r-1$.
The $2r$-dimensional representation is ($r \ge 2$)
\begin{gather}
  E_{\alpha_0} = e_{2r,2} + e_{2r-1,1} , \quad E_{\alpha_i} = e_{i,i+1} + e_{2r-i,2r+1-i} , \quad E_{\alpha_r} = e_{r,r+1} .
\end{gather}

\item[$D_{r+1}^{(2)}$:]

The Coxeter number $h=2r+2$.
The $(2r+2)$-dimensional representation is ($r \ge 2$)
\begin{equation}
\begin{gathered}
  E_{\alpha_0} = \sqrt{2}(e_{r+2,1} + e_{2r+2,r+2}) , \quad   E_{\alpha_i} = e_{i,i+1} + e_{2r+2-i,2r+3-i} , \\
  E_{\alpha_r} = \sqrt{2}(e_{r,r+1} + e_{r+1,r+3}) .
\end{gathered}
\end{equation}

\item[$A_{2r}^{(2)}$:]
The Coxeter number $h=2r$.
The $(2r+1)$-dimensional representation is ($r \ge 1$)
\begin{gather}
  E_{\alpha_0} = e_{2r+1,1} , \quad E_{\alpha_i} = e_{i,i+1} + e_{2r+1-i,2r+2-i} , \quad E_{\alpha_r} = \sqrt{2}(e_{r,r+1}+e_{r+1,r+2} .
\end{gather}

\item[$C_r^{(1)}$:]
The Coxeter number $h=2r-1$.
The $2r$-dimensional representation is ($r \ge 1$)
\begin{gather}
  E_{\alpha_0} = e_{2r,1}, \quad E_{\alpha_i} = e_{i,i+1}+e_{2r-i,2r+1-i}, \quad E_{\alpha_r} = e_{r,r+1}.
\end{gather}
\end{description}

\section{The Recursive method to $D_{r+1}^{(2)}$ pseudo-ODE }\label{Appe.B}
Due to the existence of the operator $\partial^{-1}_{z}$ in \eqref{eq: Cr ode}, the conventional WKB method to Eq. \eqref{eq: WKB expand} cannot be applied. However, it is possible to solve it in the matrix form. First, make the following WKB ansatz:
\begin{equation}\label{eq:vec_psi}
    \Psi(z,\epsilon)=(\exp(\frac{1}{\epsilon}\int^{z}P_{1}(z',\epsilon)dz'),\dots,\exp(\frac{1}{\epsilon}\int^{z} P_{2r+2}(z',\epsilon)dz'))^{\text{T}},
\end{equation}
where the functions $P_{i}(z,\epsilon)=\sum_{n=0}^{\infty}\epsilon^{n}P_{i(n)}(z)$. Substitute $\Psi$ into the linear problem. It can be shown that for $ 1\leq i\leq r\;\text{and}\; r+3\leq i\leq 2r+1,$
\begin{align}\label{eq:Cr ith relation}
\begin{split}
    &P_{i}(z,\epsilon)\exp(\frac{1}{\epsilon}\int^{z} P_{i}(z',\epsilon)dz')+\exp(\frac{1}{\epsilon}\int^{z} P_{i+1}(z',\epsilon)dz')=0,\\
    &P_{r+2}(z,\epsilon)\exp(\frac{1}{\epsilon}\int^{z} P_{r+2}(z',\epsilon)dz')+p(z)\exp(\frac{1}{\epsilon}\int^{z} P_{1}(z',\epsilon)dz')=0,\\
    &P_{2r+2}(z,\epsilon)\exp(\frac{1}{\epsilon}\int^{z} P_{2r+2}(z',\epsilon)dz')+p(z)\exp(\frac{1}{\epsilon}\int^{z} P_{r+2}(z',\epsilon)dz')=0.\\
\end{split}
\end{align}
By taking the derivative of Eq. \eqref{eq:Cr ith relation}, it is possible to eliminate the exponential functions and obtain the following relations.
\begin{align}\label{eq: Cr S recursion}
\begin{split}
    &P_{i+1}=P_{i}+\epsilon\frac{P_{i}'}{P_{i}},\quad 1\leq i\leq r,\quad r+3\leq i\leq 2r+1,\\
    &P_{1}=P_{r+2}+\epsilon\Big(\frac{P_{r+2}'}{P_{r+2}}-\frac{p'(z)}{p(z)}\Big),\\
    &P_{r+2}=P_{2r+2}+\epsilon\Big(\frac{P_{2r+2}'}{P_{2r+2}}-\frac{p'(z)}{p(z)}\Big).
\end{split}
\end{align}
Then sum up Eq. \eqref{eq: Cr S recursion} and expand the result perturbatively, we obtain
\begin{equation}\label{eq: Cr constraint}
    \sum_{i=1}^{2r+2}\Big(\frac{P_{i}'}{P_{i}}\Big)_{(n-1)}-2\delta_{1n}\Big(\frac{p'(z)}{p(z)}\Big)=0.
\end{equation}
Eq. \eqref{eq: Cr constraint} and Eq. \eqref{eq: Cr S recursion} can be used to find the WKB solution $\Psi(z,\epsilon)$ \eqref{eq:vec_psi}.

\section{WKB expansion for $\phi_{i}=0$ case}\label{Appe.c}
In this appendix, we consider the linear problem $\mathcal{L}_{m}\mathbf{\Psi}=0$ with the Lax operator given by Eq. \eqref{eq:holo_mlax} for $A_{2r-1}^{(2)}$, $B_{r}^{(1)}$, $D_{r}^{(1)}$, and $D_{r+1}^{(2)}$ when $\phi_i=0$.
To obtain the simplified formula for the WKB periods, the diagonal element $f(z,\epsilon)$ in Eq. \eqref{eq:diag_ele} admits the following simplification:
\begin{equation}
    f_{i}(f_{0},f_{0}',f_{0}'',\cdots)=S_{i}(S_{0},S_{0}'',\cdots)+d(*)
\end{equation}
where $S_{0}=f_{0}$, and $S_{i}$ are independent of $S_{0}'$. All the $S_{0}'$ dependent terms are absorbed into total derivatives $d(*)$. After the modification, $S_{2i+1}$ turns out to vanish and $S_{2i}(z)$ can be fixed by $S_{0}=f_{0}$ and the rank $r$ when $\phi_{i}=0$. 
From the WKB analysis of the linear problem with $\phi_i=0$,
we found the general formula for $r\leq 6$ cases and put them here. First, the general formula for $S_{2n}$ $(n\leq6)$ is given by\begin{align}
\begin{split}
    &S_{2}=a_{[2]1}(r)\frac{S_{0}''}{S_{0}^{2}},\\
    &S_{4}=a_{[4]1}(r)\frac{(S_{0}'')^2}{S_{0}^5}+a_{[4]2}(r)\frac{S_{0}^{(4)}}{S_{0}^4},\\
    &S_{6}=a_{[6]1}(r)\frac{(S_{0}'')^{3}}{S_{0}^{8}}+a_{[6]2}(r)\frac{(S_{0}^{(3)})^{2}}{S_{0}^{7}}+a_{[6]3}(r)\frac{S_{0}''S_{0}^{(4)}}{S_{0}^{7}}+a_{[6]4}(r)\frac{S_{0}^{(6)}}{S_{0}^{6}},\\
\end{split}
\end{align}
which can be applied to any affine Lie algebra in this paper. For different algebras, $a_{[n]i}(r)$ are different, but some of them are connected. We will show the concrete forms of $a_{[n]i}(r)$ for $A_{2r-1}^{(2)}$, $B_{r}^{(1)}$, $D_{r}^{(1)}$ and $D_{r+1}^{(2)}$ in this appendix. For $A_{r}^{(1)}$ type, the coefficients have been found in \cite{Ito:2021boh}.

\subsection{$D_{r}^{(1)}/B_{r-1}^{(1)}$}
As we have seen in Section \ref{sect:Sec_4}, the diagonal elements in $B_{r-1}^{(1)}$ types are the same as the ones in $D_{r}^{(1)}$ types when $\phi_{i}=0$.
\begin{align}\label{eq:Dr_coe}
    \begin{split}
        &a_{[2]1}=\frac{1}{24} r (2 r-1),\\
        &a_{[4]1}=\frac{1}{192} (r-4) r (2 r-1) (2 r+1),\\
        &a_{[4]2}=-\frac{(r-4) r (2 r-1) (2 r+1)}{1280},\\
        &a_{[6]1}=\frac{r (2 r-1) (2 r+3) \left(802 r^3-4867 r^2+7708 r+3360\right)}{165888},\\
        &a_{[6]2}=-\frac{r^2 (2 r-1) (2 r+3) \left(2 r^2+13 r-52\right)}{48384},\\
        &a_{[6]3}=-\frac{r (2 r-1) (2 r+3) \left(22 r^3-145 r^2+244 r+96\right)}{18432},\\
        &a_{[6]4}=\frac{r (2 r-1) (2 r+3) \left(22 r^3-145 r^2+244 r+96\right)}{774144}.\\
    \end{split}
\end{align}
From $a_{[4]1}$ and $a_{[4]2}$, one can easily find $f_{4}(z)$ term in $D_{4}^{(1)}$ or $B_{3}^{(1)}$ are total derivatives, which can be viewed as a special case out of the traceless condition.

\subsection{$D_{r+1}^{(2)}$}
$a_{[n]i}(r)$ for $D_{r+1}^{(2)}$ types are given by
\begin{align}
    \begin{split}
        &a_{[2]1}=\frac{1}{24} r (2 r+1),\\
        &a_{[4]1}=\frac{1}{192} r (r+4) (2 r-1) (2 r+1),\\
        &a_{[4]2}=-\frac{r (r+4) (2 r-1) (2 r+1)}{1280},\\
        &a_{[6]1}=\frac{r (2 r-3) (2 r+1) \left(802 r^3+4867 r^2+7708 r-3360\right)}{165888},\\
        &a_{[6]2}=-\frac{r^2 (2 r-3) (2 r+1) \left(2 r^2-13 r-52\right)}{48384},\\
        &a_{[6]3}=-\frac{r (2 r-3) (2 r+1) \left(22 r^3+145 r^2+244 r-96\right)}{18432},\\
        &a_{[6]4}=\frac{r (2 r-3) (2 r+1) \left(22 r^3+145 r^2+244 r-96\right)}{774144}.\\
    \end{split}
\end{align}
Compared with \eqref{eq:Dr_coe}, there exists a $r\leftrightarrow-r$ symmetry between $D_{r}^{(1)}$ and $D_{r+1}^{(2)}$.

\subsection{$A_{2r-1}^{(2)}$}
Finally, $a_{[n]i}(r)$ for $A_{2r-1}^{(2)}$ types are given by
\begin{align}
    \begin{split}
        &a_{[2]1}=\frac{1}{24} r (2 r-1),\\
        &a_{[4]1}=\frac{1}{192} r (r+1) (2 r-7) (2 r+1),\\
        &a_{[4]2}=-\frac{r (r+1) (2 r-7) (2 r+1)}{1280},\\
        &a_{[6]1}=\frac{r (r+2) (2 r+1) \left(1604 r^3-7328 r^2+6885 r+12195\right)}{165888},\\
        &a_{[6]2}=-\frac{r (r+2) (2 r+1)^2 \left(2 r^2+15 r-45\right)}{48384},\\
        &a_{[6]3}=-\frac{r (r+2) (2 r+1) \left(44 r^3-224 r^2+231 r+369\right)}{18432},\\
        &a_{[6]4}=\frac{r (r+2) (2 r+1) \left(44 r^3-224 r^2+231 r+369\right)}{774144}.\\
    \end{split}
\end{align}

\end{document}